\documentclass[11pt,a4paper]{article}
\DeclareMathAlphabet{\scr}{U}{rsfs}{m}{n}

\pdfoutput=1
\usepackage{latexsym}
\usepackage{epsfig}
\usepackage[mathscr]{eucal}
\usepackage{amsfonts}
\usepackage{amscd}
\usepackage{cite}
\usepackage{array}   
\usepackage{amssymb}
\usepackage{colordvi}
\usepackage[centertags]{amsmath}
\usepackage{enumerate}
\usepackage{graphicx}
\usepackage{booktabs}
\usepackage{hyperref}
\usepackage{enumitem}
\setlist[description]{leftmargin=2\parindent,labelindent=\parindent}
\usepackage{theorem}
\usepackage{listings}
\usepackage{multirow}
\usepackage[framemethod=TikZ]{mdframed}
\usepackage{makecell}
\usepackage[footnotesize]{caption}
\usepackage{bbm}
\usepackage[labelfont=bf,textfont={sl,bf}]{subfig}
 \usepackage{slashed}
\usepackage{xcolor}
\xdefinecolor{mygray}{gray}{0.85}
\usepackage{ulem}
\usepackage{pifont}
\newcommand{\cmark}{\ding{51}}%
\newcommand{\xmark}{\ding{55}}%
\usepackage{mathtools}

\mdfsetup{%
   middlelinecolor=black,
   middlelinewidth=0.5pt,
   roundcorner=0pt,
   innertopmargin=-3pt,
   innerbottommargin=10pt
}

\newcommand{\newc}{\newcommand}
\newc{\be}{\begin{equation}}
\newc{\ee}{\end{equation}}
\newc{\bea}{\begin{eqnarray}}
\newc{\eea}{\end{eqnarray}}
\newc{\ol}{\overline}
\newc{\bs}{\boldsymbol}
\newc{\m}{\mathcal}
\newc{\lan}{\langle}
\newc{\ra}{\rangle}
\newc{\pa}{\partial}

\newcommand{\beq}{\begin{eqnarray}} 
\newcommand{\eeq}{\end{eqnarray}} 
\newcommand{\bpmatrix}{\begin{pmatrix}}
\newcommand{\epmatrix}{\end{pmatrix}}
\newcommand{\ba}{\begin{array}}
\newcommand{\ea}{\end{array}}

\newcommand{\figref}[1]{Fig.~\ref{#1}}
\renewcommand{\eqref}[1]{Eq.~(\ref{#1})}

\newcommand{\bc}{\begin{center}}
\newcommand{\ec}{\end{center}}

\renewcommand{\ol}{\text{1l}}

\begin{document}
\title{
\vspace*{-3cm}
\phantom{h} \hfill\mbox{\small KA-TP-05-2019}
\\[2cm]
\textbf{{\texttt{ewN2HDECAY}} - A program for the Calculation of
  Electroweak One-Loop Corrections to Higgs Decays in the Next-to-Minimal Two-Higgs-Doublet
  Model Including State-of-the-Art QCD Corrections}}

\date{}
\author{Marcel Krause$^{1\,}$\footnote{E-mail:
  \texttt{marcel.krause@kit.edu}}, Margarete M\"{u}hlleitner$^{1\,}$\footnote{E-mail:
  \texttt{margarete.muehlleitner@kit.edu}}
\\[9mm]
{\small\it
$^1$Institute for Theoretical Physics, Karlsruhe Institute of Technology,} \\
{\small\it Wolfgang-Gaede-Str. 1, 76131 Karlsruhe, Germany.}\\[3mm]
}
\maketitle

\begin{abstract}
We present in this paper our new program package {\texttt{ewN2HDECAY}}
for the calculation of the partial decay widths and branching ratios of the Higgs
bosons of the Next-to-Minimal 2-Higgs Doublet Model (N2HDM). The N2HDM
is based on a general CP-conserving 2HDM which is extended by a real
scalar singlet field. The program computes the complete electroweak
one-loop corrections to all non-loop-induced two-body on-shell Higgs
boson decays in the N2HDM and combines them with the state-of-the-art
QCD corrections that are already implemented in the existing program
{\texttt{N2HDECAY}}. Most of the independent input parameters of the
electroweak sector of the N2HDM are renormalized in an on-shell
scheme. The soft-$\mathbb{Z}_2$-breaking squared mass scale $m_{12}^2$
and the vacuum expectation value $v_S$ of the $SU(2)_L$ singlet field,
however, are renormalized with $\overline{\text{MS}}$ conditions,
while for the four scalar mixing angles $\alpha _i$ ($i=1,2,3$) and
$\beta$ of the N2HDM, several different renormalization schemes are
applied. By giving out the leading-order and the loop-corrected
partial decay widths separately from the branching ratios, the program
{\texttt{ewN2HDECAY}} not only allows for 
phenomenological analyses of the N2HDM at highest precision, it can
also be used for a study of the impact of the electroweak corrections
and the remaining theoretical uncertainty due to missing higher-order
corrections based on a change of the renormalization scheme. The input
parameters are then consistently calculated with a parameter
conversion routine when switching from one renormalization scheme to
the other. The latest version of the 
program {\texttt{ewN2HDECAY}} can be downloaded from the URL
\href{https://github.com/marcel-krause/ewN2HDECAY}{https://github.com/marcel-krause/ewN2HDECAY}.  
\end{abstract}
\thispagestyle{empty}
\vfill
\newpage

\section{Introduction}
\label{sec:Introduction}
The discovery of the Higgs boson by the LHC experiments ATLAS
\cite{Aad:2012tfa} and CMS \cite{Chatrchyan:2012xdj}
has been a tremendous success for particle physics. The Standard Model
(SM)-like behaviour of the discovered Higgs boson \cite{Khachatryan:2016vau}, on
the other hand, leaves many questions open. To solve these problems,
extensions beyond the SM (BSM) are considered, which usually entail
enlarged Higgs sectors. In view of the lack of any direct experimental
sign of new physics (NP) so far, indirect searches for NP in the Higgs sector
become increasingly important. Due to the very SM-like nature
of the discovered Higgs boson and because of the similarity of
signatures predicted by different models, such searches require
sophisticated experimental techniques on the one side and
high-precision predictions by theory on the other side. This renders the
inclusion of higher-order corrections in the Higgs boson observables
indispensable. We contribute to this effort with the program code
that we present and publish here. In an earlier work
\cite{Krause:2018wmo}, we have published the code {\tt
  2HDECAY} for the computation of the electroweak (EW) corrections to
the on-shell non-loop induced Higgs decays of the 2-Higgs-Doublet
Model (2HDM). Here, we present the code {\tt ewN2HDECAY}. It computes
the EW corrections to the on-shell non-loop induced Higgs decays of
the Next-to-2HDM (N2HDM). The N2HDM is based on the CP-conserving 2HDM
extended by a real scalar singlet field. After electroweak symmetry breaking
(EWSB) the Higgs sector consists of three neutral CP-even, one neutral
CP-odd and two charged Higgs bosons. Due to its enlarged parameter space and
its fewer symmetries compared with supersymmetric models it
provides an interesting phenomenology with a variety of non-SM-like
signatures that are still compatible with current experimental
constraints
\cite{Muhlleitner:2016mzt,Muhlleitner:2017dkd,Azevedo:2018llq}.
Depending on which of its global symmetries are broken, its
phenomenology can change considerably \cite{Engeln:2018mbg}. Thus it
can feature a dark sector and extra sources of CP violation that only
exist in the dark sector \cite{Azevedo:2018fmj}.  In
\cite{Krause:2017mal}, we computed the EW corrections to the Higgs
boson decays in the N2HDM and provided a gauge-independent
renormalization of the N2HDM. For this, we had to extend our formalism
developed for the 2HDM in
\cite{Krause:2016gkg,Krause:2016oke,Krause:2016xku}\footnote{For later
  works discussing the renormalization of the 2HDM, see
  \cite{Denner:2016etu,Altenkamp:2017ldc,Altenkamp:2017kxk,Denner2018,Fox:2017hbw}. An
  improved on-shell scheme that is essentially equivalent to the
    mixing angle renormalization scheme presented by our group in
    \cite{Krause:2016gkg,Krause:2016oke,Krause:2016xku} is used in
    \cite{Kanemura:2017wtm,Kanemura:2017gbi,Kanemura:2018yai}.} to the N2HDM. 

We have implemented 10 different renormalization schemes in {\tt
  ewN2HDECAY} for the calculation of the EW corrections to the N2HDM
Higgs decays into all possible on-shell two-particle final states of the
model that are not loop-induced. The program is linked with the Fortran
code {\tt N2HDECAY} \cite{Muhlleitner:2016mzt}. Based on an extension
of the Fortran code {\tt HDECAY} \cite{DJOUADI199856,Djouadi:2018xqq},
{\tt N2HDECAY} incorporates the state-of-the-art higher-order QCD corrections to
the decays including also loop-induced and off-shell decays. We
consistently combine these corrections with our newly provided N2HDM
EW corrections, so that {\tt ewN2HDECAY} provides the N2HDM Higgs
boson decay widths at the presently highest possible level of
precision. Moreover, we separately give out the leading order (LO) and
next-to-leading order (NLO) EW-corrected decay widths so that 
studies can be performed on the importance of the relative EW corrections.
The comparison of the results for different renormalization schemes
additionally allows to estimate the remaining theoretical uncertainty
due to missing higher-order corrections. For the consistent comparison
we include in {\tt ewN2HDECAY} a routine that automatically converts
the input parameters from one renormalization scheme to another for all 10
implemented renormalization schemes. 

The development of {\texttt{ewN2HDECAY}} tightly followed the
development of {\texttt{2HDECAY}}, from which large parts of code were
adapted for the calculation of the Higgs boson decays in the
N2HDM. Due to the similarities of the two codes, similarities between
the structure of this paper and the manual of {\texttt{2HDECAY}},
{\it cf.}\,Ref.\,\cite{Krause:2018wmo}, are intentional. 
We still keep the description of {\tt
  ewN2HDECAY} here as short as possible while at the same time taking
care to remain self-contained. For additional details, we refer to
\cite{Krause:2018wmo} where appropriate.

The program {\texttt{ewN2HDECAY}} was developed and tested under 
{\texttt{Windows 10}}, {\texttt{openSUSE Leap 15.0}} and
{\texttt{macOS Sierra 10.12}}. In order to compile and run the
program, an up-to-date version of 
{\texttt{Python 2}} or {\texttt{Python 3}} (tested with versions
{\texttt{2.7.14}} and {\texttt{3.5.0}}), the {\texttt{FORTRAN}}
compiler {\texttt{gfortran}} and the {\texttt{GNU C}} compilers
{\texttt{gcc}} (tested for compatibility with versions
{\texttt{6.4.0}} and {\texttt{7.3.1}}) and {\texttt{g++}} are required. The latest
version of the package can be downloaded from 
\begin{center}
	\href{https://github.com/marcel-krause/ewN2HDECAY}{https://github.com/marcel-krause/ewN2HDECAY} ~.
\end{center}

The paper is organized as follows. In the subsequent
Sec.\,\ref{sec:EWQCDN2HDMMain}, we briefly introduce the N2HDM and its
particle content, focusing solely on the differences with respect to
the 2HDM due to its extended scalar sector. Moreover, we introduce the
relevant input parameters and set our notation. We briefly present the
counterterms of the extended scalar sector that are needed for the
computation of the EW corrections as well as changes in the
counterterms with respect to the 2HDM. A full list of the decays
implemented in {\texttt{ewN2HDECAY}} is presented, the link to {\tt
  N2HDECAY} is described and the parameter
conversion is discussed. In Sec.\,\ref{sec:programDescriptionMain}, we
introduce the program {\texttt{ewN2HDECAY}}, describe its structure in
detail and provide an installation and usage guide. Moreover, we
briefly describe the required format of the input and output file and
the meaning of the input parameters. We complete the paper with a
short summary of our work in Sec.\,\ref{sec:summary}. As a useful
reference for the user, we print the exemplary input and output files
which are included in the {\texttt{ewN2HDECAY}} repository, in
Appendices \ref{sec:AppendixInputFile} and
\ref{sec:AppendixOutputFile}, respectively. 

\section{One-Loop Electroweak and QCD Corrections in the N2HDM}
\label{sec:EWQCDN2HDMMain}
After a brief introduction of the N2HDM where we set up our notation
and list the input parameters used for the calculation we present the
renormalization of the EW sector of the N2HDM that we apply in the
computation of the EW one-loop corrections to the partial decay widths
of the neutral N2HDM Higgs bosons. We shortly describe the
computation of these partial decay widths and explain
how the EW-corrected partial decay widths are combined with the
state-of-the-art QCD corrections already implemented in the code
{\texttt{N2HDECAY}}. 
 
\subsection{Introduction of the N2HDM}
\label{sec:setupOfModel}
We consider a general CP-conserving N2HDM which,  
in comparison to the 2HDM, is
extended by adding a real $SU(2)_L$ singlet field $\Phi _S$ with
hypercharge $Y=0$. Together with the two complex $SU(2)_L$
doublets $\Phi _1$ and $\Phi _2$ with hypercharges $Y=+1$, it builds the
scalar sector of the model. 
Through EWSB, the two doublet and the singlet fields 
develop non-negative real vacuum expectation values (VEVs) $v_1$, $v_2$ and
$v_S$, which in general are non-vanishing. The doublet and singlet
fields can be expanded around these VEVs as 
\begin{equation}
\Phi _1 = \begin{pmatrix} \omega ^\pm _1 \\ \frac{v_1 + \rho _1 + i \eta
    _1 }{\sqrt{2}} \end{pmatrix} ~,~~~ \Phi _2 = \begin{pmatrix}
  \omega ^\pm _2 \\ \frac{v_2 + \rho _2 + i \eta
    _2}{\sqrt{2}} \end{pmatrix} ~,~~~ \Phi _S = v_S + \rho _S ~, 
\label{eq:vevexpansion}
\end{equation} 
where $\rho _i$ and $\rho _S$ are three CP-even fields, $\eta _i$ are two CP-odd fields and $\omega ^\pm _i$ are two electromagnetically charged fields ($i=1,2$). The two VEVs of the doublets are connected to the SM VEV $v$ via the relation
\begin{equation}
v_1^2 + v_2^2 = v^2 \approx (246.22~\text{GeV})^2 ~.
\label{eq:vevRelations} 
\end{equation}
The ratio of the two VEVs defines the characteristic parameter $\beta$ given by
\begin{equation}
	\tan \beta = \frac{v_2}{v_1} ~.
\label{eq:definitionTanBeta}
\end{equation}

The EW part of the Lagrangian relevant for our computation of the EW
one-loop corrections is given by
\begin{equation}
	\mathcal{L} ^\text{EW}_\text{N2HDM} = \mathcal{L} _\text{YM} + \mathcal{L} _\text{F} + \mathcal{L} _\text{S} + \mathcal{L} _\text{Yuk} + {\mathcal{L}} _\text{GF} + \mathcal{L} _\text{FP} ~.
\label{eq:electroweakLagrangian}
\end{equation}
The Yang-Mills Lagrangian $\mathcal{L} _\text{YM}$ and the fermion
Lagrangian $\mathcal{L} _\text{F}$ are analogous to the SM. Their
explicit forms are presented \textit{e.g.}\,in \cite{Peskin:1995ev,
  Denner:1991kt}. We do not present the explicit forms of the gauge-fixing
Lagrangian ${\mathcal{L}} _\text{GF}$ and Fadeev-Popov Lagrangian
$\mathcal{L}_\text{FP}$ as they are not needed in the following. We
remark, however, that we follow the same approach as in
\cite{Santos:1996vt} for the 2HDM and apply the gauge-fixing procedure
only after the renormalization of the N2HDM is performed so that
${\mathcal{L}} _\text{GF}$ contains only renormalized fields and no
additional counterterms for the gauge-fixing terms need to be
introduced. The interaction of the Higgs bosons with the fermions is
derived from the Yukawa Lagrangian $\mathcal{L}_{\text{Yuk}}$, with the
corresponding Yukawa couplings presented \textit{e.g.}\,in
\cite{Muhlleitner:2016mzt}. 

The scalar Lagrangian with the kinetic terms of the Higgs doublets and
the scalar N2HDM potential reads
\begin{equation}
\mathcal{L} _S = \sum _{i=1}^2 (D_\mu \Phi _i)^\dagger (D^\mu \Phi _i) + (\partial _\mu \Phi _S) (\partial ^\mu \Phi _S) - V_\text{N2HDM} ~,
\label{eq:scalarLagrangian}
\end{equation}
where $D_\mu$ denotes the covariant derivative
\begin{equation}
D_\mu = \partial _\mu + \frac{i}{2} g \sum _{a=1}^3 \sigma ^a W_\mu ^a
+ \frac{i}{2} g{'} B_\mu  \;,
\end{equation}
with the gauge couplings $g$ and $g{'}$ and the corresponding gauge
boson fields $W_\mu ^a$ and $B_\mu$ of the $SU(2)_L$ and $U(1)_Y$,
respectively. The generators of the $SU(2)_L$ gauge group are given by
the Pauli matrices $\sigma ^a$. The scalar potential of the
CP-conserving N2HDM is given by 
\begin{equation}
\begin{split}
V_\text{N2HDM} =&~ \frac{1}{2} m_S^2 \Phi _S^2 + \frac{1}{8} \lambda _6 \Phi _S^4 + \frac{1}{2} \lambda _7 \left( \Phi _1 ^\dagger \Phi _1 \right) \Phi _S^2 + \frac{1}{2} \lambda _8 \left( \Phi _2 ^\dagger \Phi _2 \right) \Phi _S^2 + V_\text{2HDM} ~,
\end{split}
\label{eq:scalarPotential}
\end{equation}
where $V_\text{2HDM}$ denotes the scalar potential of the CP-conserving 2HDM, as given by \cite{Branco:2011iw}
\begin{equation}
\begin{split}
V_\text{2HDM} =&~ m_{11}^2 \left| \Phi _1 \right| ^2 + m_{22}^2 \left|
  \Phi _2 \right| ^2 - m_{12}^2 \left( \Phi _1 ^\dagger \Phi _2 +
  \textit{h.c.} \right) + \frac{\lambda _1}{2} \left( \Phi _1^\dagger
  \Phi _1 \right) ^2 + \frac{\lambda _2}{2} \left( \Phi _2^\dagger
  \Phi _2 \right) ^2 \\ 
&+ \lambda _3 \left( \Phi _1^\dagger \Phi _1 \right) \left( \Phi
  _2^\dagger \Phi _2 \right) + \lambda _4 \left( \Phi _1^\dagger \Phi
  _2 \right) \left( \Phi _2^\dagger \Phi _1 \right) + \frac{\lambda
  _5}{2} \left[ \left( \Phi _1^\dagger \Phi _2 \right) ^2 +
  \textit{h.c.} \right] ~. 
\end{split}
\label{eq:scalarPotential2HDM}
\end{equation}
It is obtained by imposing two $\mathbb{Z}_2$ symmetries on the scalar
potential, under one of which, $\mathbb{Z}_2$, 
\beq
\Phi_1 \to \Phi_1 \;, \quad \Phi_2 \to -\Phi_2 \;, \quad \Phi_S \to
\Phi_S \;. \label{eq:z2one}
\eeq
It is the trivial generalisation of the usual 2HDM $\mathbb{Z}_2$
symmetry (guaranteeing the absence of tree-level flavour-changing
neutral currents (FCNCs) when extended to the Yukawa sector) and explicitly
broken by the term proportional to $m_{12}^2$. The second symmetry,
$\mathbb{Z}'_2$, under which 
\beq
\Phi_1 \to \Phi_1 \;, \quad \Phi_2 \to \Phi_2 \;, \quad \Phi_S \to
-\Phi_S \;,
\eeq
is not explicitly broken. 
The N2HDM potential contains twelve real-valued parameters, four mass
parameters $m_{11}$, $m_{22}$, $m_{12}$ and $m_S$ and eight
dimensionless coupling constants $\lambda _i$ ($i=1,...,8$). For
later convenience, we define 
\begin{equation}
	\lambda _{345} \equiv \lambda _3 + \lambda _4 + \lambda _5 ~.
\end{equation}

Inserting the expansion of the scalar doublet and singlet fields,
\eqref{eq:vevexpansion}, into the scalar potential of the N2HDM yields 
\begin{equation}
	V_\text{N2HDM} = \frac{1}{2} \left( \rho _1 ~~ \rho _2 ~~ \rho
          _S \right) M_\rho ^2 \begin{pmatrix} \rho _1 \\ \rho _2 \\
          \rho _S \end{pmatrix} + T_1 \rho _1 + T_2 \rho _2 + T_S \rho
        _S + ~~ \cdots ~, \label{eq:vpotn2hdm}
\end{equation}
where $M_\rho ^2$ denotes the $3\times 3$ mass matrix in the CP-even
scalar sector and $T_1$, $T_2$ and $T_S$ the three tadpole terms. We
demand that the VEVs represent the minimum of the potential by
applying the minimum condition 
\begin{equation}
\frac{\partial V_\text{N2HDM}}{\partial \Phi _i} \Bigg| _{\left\langle \Phi _j \right\rangle} = 0~,
\end{equation}
leading at tree level to the vanishing of the three tadpole terms,
\begin{equation}
T_1 = T_2 = T_S = 0 ~~ \text{(at tree level)} ~,
\end{equation}
which in terms of the Higgs potential parameters read
\begin{align}
\frac{T_1}{v_1} &\equiv m_{11}^2 - m_{12}^2 \frac{v_2}{v_1} + \frac{v_1^2 \lambda _1}{2} + \frac{v_2 ^2 \lambda _{345}}{2} + \frac{v_S ^2 \lambda _7}{2}  \label{eq:tadpoleCondition1} \\
\frac{T_2}{v_2} &\equiv m_{22}^2 - m_{12}^2 \frac{v_1}{v_2} + \frac{v_2^2 \lambda _2}{2} + \frac{v_1 ^2 \lambda _{345}}{2} + \frac{v_S ^2 \lambda _8}{2}  \label{eq:tadpoleCondition2} \\
\frac{T_S}{v_S} &\equiv m_S^2 + \frac{v_1 ^2 \lambda _7}{2} + \frac{v_2 ^2 \lambda _8}{2} + \frac{v_S ^2 \lambda _6}{2} ~. \label{eq:tadpoleCondition3}
\end{align}
These conditions can be used to replace $m_{11}^2$, $m_{22}^2$ and $m_S^2$
in favor of the three tadpole terms. The mass matrix of the CP-even scalar fields is given
by  
\begin{align}
M_\rho ^2 &\equiv \begin{pmatrix}
m_{12}^2 \frac{v_2}{v_1} + \lambda _1 v_1^2 & -m_{12}^2 + \lambda _{345} v_1 v_2 & \lambda _7 v_1 v_S \\ -m_{12}^2 + \lambda _{345} v_1 v_2 & m_{12}^2 \frac{v_1}{v_2} + \lambda _2 v_2^2 & \lambda _8 v_2 v_S \\ \lambda _7 v_1 v_S & \lambda _8 v_2 v_S & \lambda _6 v_S^2
\end{pmatrix} + \begin{pmatrix}
\frac{T_1}{v_1} & 0 &0 \\ 0 & \frac{T_2}{v_2} & 0 \\0 & 0 & \frac{T_S}{v_S}
\end{pmatrix} ~. \label{eq:massMatrices1}
\end{align}
By introducing three mixing angles $\alpha _i$ ($i=1,2,3$) defined in the range
\begin{equation}
	- \frac{\pi}{2} \leq \alpha _i < \frac{\pi}{2} ~,
\end{equation}
the mass matrix can be diagonalised by means of the orthogonal matrix
$R$ parametrised as\footnote{Throughout this paper, we use the
  short-hand notation $s_x \equiv \sin (x)$, $c_x \equiv \cos (x)$ and
  $t_x \equiv \tan (x)$.} 
\begin{equation}
	R = \begin{pmatrix}
	c_{\alpha _1} c_{\alpha _2} & s_{\alpha _1} c_{\alpha _2} & s_{\alpha _2} \\
	- \left( c_{\alpha _1} s_{\alpha _2} s_{\alpha _3} + s_{\alpha _1} c_{\alpha _3} \right) & c_{\alpha _1} c_{\alpha _3} - s_{\alpha _1} s_{\alpha _2} s_{\alpha _3} & c_{\alpha _2} s_{\alpha _3} \\
	-c_{\alpha _1} s_{\alpha _2} c_{\alpha _3} + s_{\alpha _1} s_{\alpha _3} & - \left( c_{\alpha _1} s_{\alpha _3} + s_{\alpha _1} s_{\alpha _2} c_{\alpha _3} \right) & c_{\alpha _2} c_{\alpha _3}
	\end{pmatrix} ~,
\end{equation}
which transforms the CP-even interaction fields into the mass
eigenstates $H_i$ ($i=1,2,3$) 
\begin{equation}
	\begin{pmatrix} H_1 \\ H_2 \\ H_3 \end{pmatrix} =  R \begin{pmatrix} \rho _1 \\ \rho _2 \\ \rho _3 \end{pmatrix} ~. \label{eq:rotationCPeven}
\end{equation}
This transformation yields the diagonalised mass matrix
\begin{equation}
	D ^2 _\rho \equiv R M^2 _\rho R^T \equiv \text{diag} \left( m_{H_1}^2 , m_{H_2}^2 , m_{H_3}^2 \right) ~, \label{eq:diagonalizationMassMatrix}
\end{equation}
where we demand the three CP-even Higgs bosons $H_i$ to be
ordered by ascending mass,
\begin{equation}
	m_{H_1} < m_{H_2} < m_{H_3} ~.
\end{equation}
The CP-odd and charged mass matrices, not shown explicitly in
Eq.~(\ref{eq:vpotn2hdm}) are equivalent to the ones in the 
2HDM and are diagonalised by the two mixing angles $\beta _\eta$ and
$\beta _{\omega ^\pm}$, respectively, which, at tree level, coincide with the angle
$\beta$ defined in \eqref{eq:definitionTanBeta}. The rotation of the
corresponding fields to the mass basis yields the CP-odd and charged
Higgs bosons $A$ and $H^\pm$ with masses $m_A$ and $m_{H^\pm}$,
respectively, as well as the massless CP-odd and charged Goldstone
bosons $G^0$ and $G^\pm$. The quartic couplings $\lambda _i$
($i=1,...,8$) of the N2HDM potential can be written in terms of the
parameters of the mass basis as \cite{Muhlleitner:2016mzt} 
\begin{align}
\lambda _1 &= \frac{1}{v^2 c_\beta ^2} \left[ \sum _{i=1}^3 m_{H_i}^2 R_{i1}^2 - \frac{s_\beta}{c_\beta} m_{12}^2 \right]  \label{eq:parameterTransformationMassToInteraction1} \\
\lambda _2 &= \frac{1}{v^2 s_\beta ^2} \left[ \sum _{i=1}^3 m_{H_i}^2 R_{i2}^2 - \frac{c_\beta}{s_\beta} m_{12}^2 \right]  \\
\lambda _3 &= \frac{1}{v^2} \left[ \frac{1}{s_\beta c_\beta} \sum _{i=1}^3 m_{H_i}^2 R_{i1} R_{i2} + 2m_{H^\pm}^2 - \frac{1}{s_\beta c_\beta} m_{12}^2 \right]  \\
\lambda _4 &= \frac{1}{v^2} \left[ \frac{m_{12}^2}{s_\beta c_\beta} + m_A^2 - 2m_{H^\pm}^2 \right]  \\
\lambda _5 &= \frac{1}{v^2} \left[ \frac{m_{12}^2}{s_\beta c_\beta} - m_A^2 \right]  \\
\lambda _6 &= \frac{1}{v_S^2} \sum _{i=1}^3 m_{H_i}^2 R_{i3}^2  \\
\lambda _7 &= \frac{1}{v v_S c_\beta } \sum _{i=1}^3 m_{H_i}^2 R_{i1} R_{i3}  \\
\lambda _8 &= \frac{1}{vv_S s_\beta } \sum _{i=1}^3 m_{H_i}^2 R_{i2} R_{i3} ~. \label{eq:parameterTransformationMassToInteraction8} 
\end{align}

The gauge sector of the N2HDM does not change with respect to the
SM. After EWSB we have for the masses of the physical gauge bosons,
the $W$ and $Z$ bosons and the photon,
\begin{align}
	m_W^2 &= \frac{g^2v^2}{4} \\
	m_Z^2 &= \frac{(g^2 + g{'}^2)v^2}{4} \\
	m_\gamma ^2 &= 0 \;.
\end{align}
The electromagnetic coupling $e$ in terms of the fine-structure constant
$\alpha_{\text{em}}$, which we use as independent input, reads
\begin{align}
	e &= \sqrt{4\pi \alpha _\text{em} } = \frac{g g{'}}{\sqrt{g^2
            + g{'} ^2}}
             \label{eq:electromagneticCouplingDefinition} \;.
\end{align}
Alternatively, the tree-level relation to the Fermi constant,
\begin{align}
	G_F &\equiv \frac{\sqrt{2} g^2}{8m_W^2} = \frac{\alpha _\text{em} \pi
}{\sqrt{2} m_W^2 \left( 1 - \frac{m_W^2}{m_Z^2} \right)} ~, \label{eq:definitionFermiConstant} 
\end{align} 
can be used to replace one of the parameters of the EW sector in
favour of $G_F$.

\begin{table}[tb]
\centering
  \begin{tabular}{ c c c c }
    \hline
    & $u$-type & $d$-type & leptons \\ \hline
    I &  $\Phi _2$ & $\Phi _2$ & $\Phi _2$ \\
    II & $\Phi _2$ & $\Phi _1$ & $\Phi _1$ \\
    lepton-specific & $\Phi _2$ & $\Phi _2$ & $\Phi _1$ \\
    flipped & $\Phi _2$ & $\Phi _1$ & $\Phi _2$ \\
    \hline
  \end{tabular}
    \caption{The four realisations of the Yukawa couplings of the
      $\mathbb{Z}_2$-symmetric N2HDM, depending on which Higgs doublet
      couples to which kind of fermions.}
   \label{tab:yukawaDefinitions}
\end{table}
The aforementioned softly broken $\mathbb{Z}_2$ symmetry,
Eq.~(\ref{eq:z2one}), for the two doublet fields $\Phi _1$ and $\Phi _2$,
extended to the Yukawa sector to avoid FCNCs at tree level, leads to
four types of doublet couplings to the fermion fields as in the
2HDM. They are summarised in Tab.\,\ref{tab:yukawaDefinitions}. 
The Yukawa couplings can be parametrised in terms of the Yukawa
coupling parameters $Y^{u,d,l}_{i,4}$ ($i=1,2,3$) summarised in
Tab.~\ref{tab:yukawaCouplings}, where $i$ refers the the scalar Higgs
bosons $H_i$ and 4 to the pseudoscalar $A$.  
\begin{table}[tb]
\centering
  \begin{tabular}{ c c c c c c c  }
    \hline
    N2HDM type & $Y^l_i$ & $Y^l_4$ & $Y^d_i$ & $Y^d_4$ & $Y^u_i$ & $Y^u_4$ \\ \hline
    I  & $\frac{R_{i2} }{s_\beta }$ & $\frac{1}{t_\beta}$ & $\frac{R_{i2}}{s_\beta }$ &
    $\frac{1}{t_\beta}$ & $\frac{R_{i2} }{s_\beta }$ & $\frac{1}{t_\beta}$\\
    II & $\frac{R_{i1} }{c_\beta }$ & $-t_\beta$ & $\frac{R_{i1}}{c_\beta }$ &$-t_\beta$ 
 & $\frac{R_{i2} }{s_\beta }$ & $\frac{1}{t_\beta}$\\
    lepton-specific  & $\frac{R_{i1} }{c_\beta }$ & $-t_\beta$ &
$\frac{R_{i2} }{s_\beta }$ & $\frac{1}{t_\beta}$ & $\frac{R_{i2} }{s_\beta }$ & $\frac{1}{t_\beta}$\\
    flipped  & $\frac{R_{i2} }{s_\beta }$ & $\frac{1}{t_\beta}$ &$\frac{R_{i1}}{c_\beta}$ &
$-t_\beta$ & $\frac{R_{i2} }{s_\beta }$ & $\frac{1}{t_\beta}$\\
    \hline
  \end{tabular}
    \caption{Definition of the Yukawa coupling parameters
      $Y^{u,d,l}_{i,4}$ ($i=1,2,3$) for each N2HDM type.}
   \label{tab:yukawaCouplings}
\end{table}

Apart from the parameters introduced so far, {\texttt{N2HDECAY}}
requires additional input parameters, namely the electromagnetic
fine-structure constant $\alpha _\text{em}$ in the Thomson limit for
the computation of the loop-induced decays into $\gamma\gamma$ and
$Z\gamma$, the strong coupling constant $\alpha _s$ for the 
calculation of the loop-induced decays into gluons as well as for the
state-of-the-art QCD corrections, and additionally the total decay
widths $\Gamma _W$ and $\Gamma _Z$ of the $W^\pm$ and $Z$ bosons,
respectively, for the calculation of the off-shell decays into pairs
of these gauge bosons. 
Moreover, the CKM matrix elements $V_{ij}$ are required as input for
the computation of the decays involving charged 
quark currents, as well as the quark masses $m_f$, where
$f=s,c,b,t,\mu ,\tau$ -- all other fermions are approximated to be
massless in the computation of the partial decay widths. The fermion
and gauge boson masses are defined according to the recommendations of
the LHC Higgs cross section working group \cite{Denner:2047636}. Note
that in {\texttt{N2HDECAY}} the decay widths are computed in terms of
the Fermi constant $G_F$ with the exception of the decays involving
external on-shell photons which are given in terms of
$\alpha_{\text{em}}$ in the Thomson limit. As input for our
renormalization conditions of the EW corrections, we require as input
parameters, however, the on-shell masses $m_W$ and $m_Z$ and the
electromagnetic coupling at the $Z$ boson mass scale,
$\alpha_{\text{em}}(m_Z^2)$. We will come back to this in
Sec.\,\ref{sec:connectionHDECAY}. The full set of independent input
parameters in the mass basis therefore is given by ($i=1,2,3$)
\begin{equation}
\{ G_F, \alpha _s , \Gamma _W , \Gamma _Z , \alpha _\text{em} , m_W ,
m_Z , m_f, V_{ij} , t_\beta , m_{12}^2 , v_S, \alpha _i , m_{H_i} , m_A ,
m_{H^\pm} \} ~.
\label{eq:inputSetMassBase} 
\end{equation}

For completeness, we want to mention
that the three tadpole parameters $T_1$, $T_2$ and $T_S$ are formally
independent input values, as well. However, as we explain in the
upcoming Sec.\,\ref{sec:renormalizationTadpoles}, these parameters are
either zero after renormalization or they are not present in the
theory in the first place within an alternative treatment of the
minimum conditions of the potential. In both cases, the
parameters are effectively required to vanish and hence, we do not
include them in the set of independent input parameters. 

\subsection{Renormalization}
\label{sec:renormalization2HDM}
The ultraviolet (UV) divergences appearing in the one-loop
corrections to the decay widths calculated in this work require the
renormalization of the relevant parameters entering the decay widths
at tree level. Apart from an extended CP-even scalar
sector, the N2HDM is equivalent to the 2HDM at tree level so that 
the one-loop renormalization of the N2HDM can be performed 
analogously to the 2HDM. In the following, we only present the
renormalization of the extended CP-even scalar sector of the N2HDM,
\textit{i.e.}\,we describe only the changes of the renormalization
with respect to the 2HDM. For a thorough discussion of the
renormalization of the N2HDM in general, we refer to
\cite{Krause:2017mal} and for the definition of the 2HDM-like
counterterms (CTs) as they are incorporated into
{\texttt{ewN2HDECAY}}, we refer to \cite{Krause:2016gkg,
  Krause:2016oke, Krause:2016xku, Krause:2018wmo}.\footnote{The
  renormalization of the 2HDM is also discussed in detail
  in~\cite{Denner:2016etu,Altenkamp:2017ldc} and in \cite{Denner2018},
  the renormalization of mixing angles in general and the application
  to the 2HDM in particular is
  discussed. In~\cite{Fox:2017hbw,Grimus:2018rte}, the
  gauge-independent renormalization of multi-Higgs models is described.}

Most of the N2HDM input parameters given in
\eqref{eq:inputSetMassBase} are renormalized with on-shell (OS) conditions. For
the physical masses, this leads to their
counterterms being defined as the real parts of the poles of the propagators of the
corresponding fields. All physical fields are equivalently
renormalized in an OS approach, \textit{i.e.}\,we demand that on the
mass shell, no mixing of fields with the same quantum numbers takes
place. The residues of the propagators associated to
the corresponding fields are normalized to unity so that the fields
are properly normalized as well. The soft-$\mathbb{Z}_2$-breaking
parameter $m_{12}^2$ and the singlet vacuum expectation value $v_S$
are renormalized via $\overline{\text{MS}}$ conditions. For the scalar
mixing angles $\alpha _i$ ($i=1,2,3$) and $\beta$, several different 
renormalization schemes are implemented. 

\subsubsection{Renormalization of the Tadpoles}
\label{sec:renormalizationTadpoles}
As discussed \textit{e.g.}\,in \cite{Krause:2016gkg, Krause:2016oke},
the proper renormalization of the ground state of the Higgs potential
is crucial for defining gauge-independent counterterms for the scalar
mixing angles. We have implemented two different procedures in
{\texttt{ewN2HDECAY}} that we briefly introduce\footnote{For further information,
  see \cite{Krause:2016gkg,
    Krause:2016oke,Krause:2018wmo,Krause:2017mal} where they have been
  discussed in detail.} before presenting the related renormalization conditions.

At NLO the tadpole terms are replaced by
\begin{equation}
	T_i ~ \longrightarrow ~ T_i + \delta T_i ~~~ (i=1,2,S) \;,
\end{equation}
where the $T_i$ at the right-hand side denote the renormalized tadpole
terms and $\delta T_i$ the corresponding CTs.
In the \textit{standard tadpole scheme}, used \textit{e.g.}\,in \cite{Denner:1991kt}
for the SM and in \cite{Kanemura:2004mg, Kanemura:2015mxa} for the
2HDM, the tadpole terms, which define the ground state of the Higgs
potential, are renormalized such that the potential remains at the
correct minimum at higher orders. This implies the renormalization
conditions $T_i = 0$ at one-loop level and eventually leads to the
identification of the tadpole CTs $\delta T_i$ with the corresponding
genuine one-loop tadpole diagrams $T^\text{loop}_i$. Denoting the 
counterterms and one-loop tadpole diagrams in the mass basis by 
$\delta T_{H_i}$ and $T^\text{loop}_{H_i}$, respectively, we have
\begin{equation}
i\delta T_{H_i} = iT^\text{loop}_{H_i} = \mathord{ \left(\rule{0cm}{30px}\right. \vcenter{
    \hbox{ \includegraphics[height=57px , trim = 18.2mm 12mm 17.3mm
      10mm, clip]{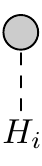} } }
  \left.\rule{0cm}{30px}\right)  } ~~~ (i=1,2,3) ~.
\label{eq:tadpoleCountertermDefinition}
\end{equation}
The tadpole CTs $\delta T_{H_i}$ in the mass basis are related to
those in the gauge basis, $\delta T_i$, through the rotation with the
mixing matrix $R$ defined in \eqref{eq:rotationCPeven}, 
\begin{align}
	\delta T_1 &= \sum _{j=1}^3 R_{j1} \delta T_{H_j} \\
	\delta T_2 &= \sum _{j=1}^3 R_{j2} \delta T_{H_j} \\	
	\delta T_S &= \sum _{j=1}^3 R_{j3} \delta T_{H_j} ~.	
\end{align}
The tadpole terms appear in the diagonal entries of all scalar mass
matrices, {\it cf.}~\textit{e.g.}\,\eqref{eq:massMatrices1}. After their
diagonalisation, {\it cf.}\,\eqref{eq:diagonalizationMassMatrix}, this
leads to twelve different tadpole CTs at NLO, given by
\begin{mdframed}[frametitle={Renormalization of the tadpoles (standard scheme)},frametitlerule=true,frametitlebackgroundcolor=black!14,frametitlerulewidth=0.6pt]\begin{align}
\delta T_{H_iH_j} &= R_{i1} R_{j1} \frac{\delta T_1}{v_1} + R_{i2} R_{j2} \frac{\delta T_2}{v_2} + R_{i3} R_{j3} \frac{\delta T_S}{v_S} ~~~ (i,j=1,2,3)  \label{eq:RenormalizationRadpolesTadpoleCountertermDeltaTH0H0ExplicitForm} \\
\delta T_{G^0G^0} &= c_\beta \frac{\delta T_1 }{v} + s _\beta \frac{ \delta T_2 }{v}  \label{eq:RenormalizationRadpolesTadpoleCountertermDeltaTG0G0ExplicitForm} \\
\delta T_{G^0A} &= -s_\beta \frac{\delta T_1 }{v} + c _\beta \frac{ \delta T_2 }{v}  \\
\delta T_{AA} &= \frac{s_\beta ^2}{c_\beta} \frac{\delta T_1 }{v} + \frac{c_\beta ^2}{s_\beta} \frac{ \delta T_2 }{v}  \label{eq:RenormalizationRadpolesTadpoleCountertermDeltaTA0A0ExplicitForm} \\
\delta T_{G^\pm G^\pm } &= c_\beta \frac{\delta T_1 }{v} + s _\beta \frac{ \delta T_2 }{v}  \\ 
\delta T_{G^\pm H^\pm } &= -s_\beta \frac{\delta T_1 }{v} + c _\beta \frac{ \delta T_2 }{v}  \\ 
\delta T_{H^\pm H^\pm } &= \frac{s_\beta ^2}{c_\beta} \frac{\delta T_1 }{v} + \frac{c_\beta ^2}{s_\beta} \frac{ \delta T_2 }{v} ~. 
\label{eq:RenormalizationRadpolesTadpoleCountertermDeltaTHpHpExplicitForm}
\end{align}\end{mdframed}
The renormalization conditions of \eqref{eq:tadpoleCountertermDefinition} 
imply that no tadpole diagrams have to be taken into account in the calculation of partial
decay widths at one-loop level apart from the CTs of the scalar mass
matrices, in which the tadpole terms explicitly appear and as a 
consequence, the tadpole CTs in the mass basis from
Eqs.\,(\ref{eq:RenormalizationRadpolesTadpoleCountertermDeltaTH0H0ExplicitForm})-(\ref{eq:RenormalizationRadpolesTadpoleCountertermDeltaTHpHpExplicitForm})
appear in the off-diagonal scalar wave function renormalization counterterms
(WFRCs) and mass CTs, {\it cf.}~Sec.\,\ref{sec:renormalizationScalarSector}. 

With the VEVs in the standard tadpole scheme being defined as the
minimum of the loop-corrected gauge-dependent potential, they are consequently gauge-dependent as well, as are all other CTs defined through this
minimum like {\it e.g.}~the mass CTs of the Higgs and gauge
bosons. While this is no problem as long as all gauge dependences in
the calculation of the loop-corrected decay widths cancel, an improper
renormalization condition for the mixing angle CTs in the N2HDM (and
also the 2HDM) can lead to residual gauge dependences that spoil the
overall gauge independence of the partial decay widths as discussed
in more detail in Sec.\,\ref{sec:renormalizationMixingAngles}. 
\begin{figure}[t!]
\centering
\includegraphics[width=\linewidth, trim=2.2cm 2.2cm 0cm 2.2cm, clip]{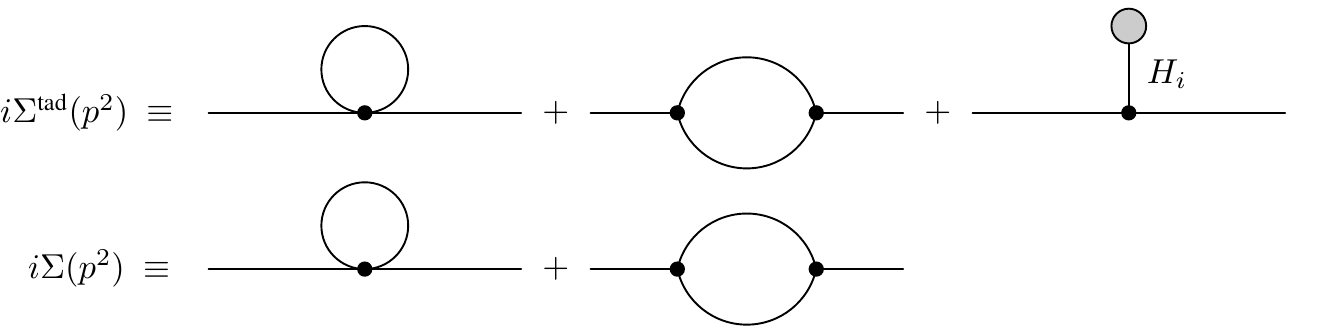}
\caption{The topologies of the generic self-energies $\Sigma$ and
  $\Sigma^\text{tad}$ as functions of the squared external momentum
  $p^2$ differ by explicit one-particle-reducible tadpole
  contributions, mediated by the three CP-even Higgs bosons $H_i$
  ($i=1,2,3$).}  
\label{fig:definitionOfSelfenergies}
\end{figure} 

In the framework of the \textit{alternative (FJ) tadpole scheme},
based on the work by J. Fleischer and F. Jegerlehner in the SM, {\it
  cf.}\,Ref.\,\cite{PhysRevD.23.2001}, and applied to the 2HDM for the
first time in Refs.\,\cite{Krause:2016gkg, Krause:2016oke}, the
so-called \textit{proper} VEVs are defined as the true ground state of
the Higgs potential, \textit{i.e.}\,as the renormalized all-order VEVs
of the Higgs fields. In this alternative framework, the VEVs are
defined through the gauge-independent tree-level potential instead of
the loop-corrected gauge-dependent one so that they become manifestly
gauge-independent quantities and thereby also the mass (matrix) CTs
become manifestly gauge-independent. 
Being the fundamental quantities in the alternative tadpole schemes,
the VEVs - instead of the tadpole terms - are shifted as
\begin{equation}
	v_i ~ \rightarrow ~ v_i + \delta v_i ~,~~~ (i=1,2,S) \label{eq:vevShifts}
\end{equation}
where the CTs $\delta v_i$ ($i=1,2,S$) are fixed by demanding that the
renormalized VEVs represent the proper tree-level minima of the Higgs
potential to all orders. This renormalization condition connects the
CTs of the VEVs with explicit tadpole diagrams as follows: 
\begin{align}
\begin{pmatrix} \delta v_1 \\ \delta v_2 \\ \delta v_S \end{pmatrix} =
  R^T \begin{pmatrix} \frac{T^\text{loop}_{H_1}}{m_{H_1}^2} \\
    \frac{T^\text{loop}_{H_2}}{m_{H_2}^2} \\
    \frac{T^\text{loop}_{H_3}}{m_{H_3}^2} \end{pmatrix} ~. 
\label{eq:vevCountertermDefinition} 
\end{align}

The renormalization of the minimum of the potential in the alternative
tadpole scheme implies a shift of the VEVs by additional tadpole
contributions. These have to be considered everywhere in the N2HDM
where the three VEVs $v_1$, $v_2$ and $v_S$ appear. Consequently, all
self-energies $\Sigma ^\text{tad}$ used for the definition of the CTs
acquire additional one-particle-reducible tadpole contributions
compared to the usual one-particle-irreducible self-energies $\Sigma$,
{\it cf.}~\figref{fig:definitionOfSelfenergies}. In the calculation of
the one-loop vertex corrections the tadpole diagrams have to be taken
into account as well, so that the alternative tadpole scheme is
characterized by the following conditions: 

\begin{mdframed}[frametitle={Renormalization of the tadpoles (alternative FJ scheme)},frametitlerule=true,frametitlebackgroundcolor=black!14,frametitlerulewidth=0.6pt]\begin{align}
\delta T_{ij} &= 0 ~, \\
\Sigma (p^2) ~ &\rightarrow ~\Sigma ^\text{tad} (p^2) ~, \\
\text{Tadpole diagrams have to be }&\text{considered in the vertex
                                     corrections.} 
\nonumber
\end{align}\end{mdframed}

\subsubsection{Renormalization of the Gauge Sector}
\label{sec:renormalizationGaugeSector}
The gauge sector of the N2HDM does not differ from that of
the 2HDM. We implement the same renormalization conditions of the
gauge sector as presented in \cite{Krause:2018wmo} for the
2HDM. Formally, all CTs are the same as stated there, they only differ
in the particle content of the two-point functions used for the definition of the CTs
of the N2HDM due to the extended scalar sector. We therefore do not
state them explicitly here and refer to \cite{Krause:2018wmo} for details. 

\subsubsection{Renormalization of the Scalar Sector}
\label{sec:renormalizationScalarSector}
The fields and masses of the CP-even scalar particles are shifted at
one-loop level as
\begin{align}
m_{H_i}^2 ~ &\rightarrow ~ m_{H_i}^2 + \delta m_{H_i}^2 ~~~ (i=1,2,3) ~, \\
\begin{pmatrix}  H_1 \\ H_2 \\ H_3 \end{pmatrix} ~ &\rightarrow ~ \begin{pmatrix} 1 + \frac{\delta Z _{{H_1} {H_1}}}{2} & \frac{\delta Z _{{H_1} {H_2}}}{2} & \frac{\delta Z _{{H_1} {H_3}}}{2}  \\ \frac{\delta Z _{{H_2} {H_1}}}{2} & 1 + \frac{\delta Z _{{H_2} {H_2}}}{2} & \frac{\delta Z _{{H_2} {H_3}}}{2} \\ \frac{\delta Z _{{H_3} {H_1}}}{2} & \frac{\delta Z _{{H_3} {H_2}}}{2} & 1 + \frac{\delta Z _{{H_3} {H_3}}}{2} \end{pmatrix} \renewcommand*{\arraystretch}{1.1} \begin{pmatrix}  H_1 \\ H_2 \\ H_3 \end{pmatrix} ~. \label{RenormalizationOnShellLabelSectionScalarSectorFieldRenormalizationConstantsCPEvenHiggses} 
\end{align}
All scalar fields are renormalized through OS conditions implying the
following CT and WFRC definitions,
\begin{mdframed}[frametitle={Renormalization of the scalar sector (standard scheme)},frametitlerule=true,frametitlebackgroundcolor=black!14,frametitlerulewidth=0.6pt,nobreak=true]\begin{align}
\delta Z_{H_iH_j} &= \frac{2}{m_{H_i}^2 - m_{H_j}^2} \textrm{Re} \Big[ \Sigma _{H_iH_j} (m_{H_j}^2) - \delta T_{H_iH_j} \Big] ~~~ (i\neq j) ~, \label{RenormalizationScalarFieldsMassesExplicitFormWaveFunctionRenormalizationConstantH0h0} \\
\delta m_{H_i}^2 &= \textrm{Re} \Big[ \Sigma _{H_iH_i} (m_{H_i}^2) - \delta T_{H_iH_i} \Big] ~,
\end{align}\end{mdframed}
\begin{mdframed}[frametitle={Renormalization of the scalar sector (alternative FJ  scheme)},frametitlerule=true,frametitlebackgroundcolor=black!14,frametitlerulewidth=0.6pt,nobreak=true]\begin{align}
\delta Z_{H_iH_j} &= \frac{2}{m_{H_i}^2 - m_{H_j}^2} \textrm{Re} \Big[ \Sigma ^\textrm{tad} _{H_iH_j} (m_{H_j}^2) \Big] ~~~ (i\neq j) ~, \label{RenormalizationScalarFieldsMassesExplicitFormWaveFunctionRenormalizationConstantH0h0Alt} \\
\delta m_{H_i}^2 &= \textrm{Re} \Big[ \Sigma ^\textrm{tad} _{H_iH_i} (m_{H_i}^2) \Big] ~, \label{RenormalizationScalarFieldsMassesExplicitFormMassCountertermH0}
\end{align}\end{mdframed}
\begin{mdframed}[frametitle={Renormalization of the scalar sector
    (standard and alternative FJ scheme)},frametitlerule=true,frametitlebackgroundcolor=black!14,frametitlerulewidth=0.6pt,nobreak=true]\begin{align}
\delta Z_{H_iH_i} &= - \textrm{Re} \left[ \frac{\partial \Sigma _{H_i H_i } \left( p^2 \right) }{\partial p^2 } \right] _{p^2 = m_{H_i} ^2} ~, \label{RenormalizationScalarFieldsMassesExplicitFormWaveFunctionRenormalizationConstantH0H0} 
\end{align}\end{mdframed}
where the tadpole CTs in the standard scheme are given in
\eqref{eq:RenormalizationRadpolesTadpoleCountertermDeltaTH0H0ExplicitForm}. The
renormalization of the other scalar particles of the N2HDM, namely of the
CP-odd and charged Higgs bosons $A$ and $H^\pm$ and the Goldstone
bosons $G^0$ and $G^\pm$, is equivalent to their renormalization
within the 2HDM and the CTs are implemented as presented in
\cite{Krause:2018wmo}. 

\subsubsection{Renormalization of the Scalar Mixing Angles}
\label{sec:renormalizationMixingAngles}
The bare mixing angles $\alpha _i$ ($i=1,2,3$) and $\beta$ are
promoted to one-loop order by introducing their CTs $\delta \alpha _i$
($i=1,2,3$) and $\delta \beta$ according to 
\begin{align}
	\alpha _i ~ &\rightarrow ~ \alpha _i + \delta \alpha _i ~~~~ (i=1,2,3) \label{eq:mixingAngleAlphaHO} \\
	\beta ~ &\rightarrow ~ \beta + \delta \beta \;, \label{eq:mixingAngleBetaHO}
\end{align}
where the mixing angles on the right-hand side are the renormalized
ones. In the following, we briefly describe the different
renormalization schemes for the scalar mixing angles that are
implemented in {\texttt{ewN2HDECAY}}. For a detailed derivation and
description of these schemes, we refer to \cite{Krause:2017mal}. \\ 

\textbf{$\overline{\text{MS}}$ scheme.} It has been shown in
Refs.\,\cite{Lorenz:2015, Krause:2016gkg} for the 2HDM that the
one-loop corrected partial decay widths can become very large when the
mixing angles are renormalized in the $\overline{\text{MS}}$ scheme
and while this can also be expected in the N2HDM, 
we nevertheless implemented the $\overline{\text{MS}}$ scheme for the
scalar mixing angles as a reference scheme. By imposing
$\overline{\text{MS}}$ conditions, only UV-divergent parts are
assigned to the four mixing angle CTs, but no finite parts
$\left. \delta \alpha _i \right| _\text{fin}$ ($i=1,2,3$) and
$\left. \delta \beta \right| _\text{fin}$. After having checked
explicitly for UV finiteness of all partial decay widths, the mixing
angle CTs in the $\overline{\text{MS}}$ scheme are implemented in
{\texttt{ewN2HDECAY}} by effectively setting them to zero: 
\begin{mdframed}[frametitle={ Renormalization of $\delta \alpha _i$ and $\delta \beta$: $\overline{\text{MS}}$ scheme (both schemes) },frametitlerule=true,frametitlebackgroundcolor=black!14,frametitlerulewidth=0.6pt,nobreak=true]\begin{align}
\left. \delta \alpha _i \right| _\text{fin} &= 0 ~~~~ (i=1,2,3) \\
\left. \delta \beta \right| _\text{fin} &= 0 
\end{align}\end{mdframed}
As a consequence of the $\overline{\text{MS}}$ renormalization, the
CTs $\delta \alpha _i$ and $\delta \beta$ become a function of the
renormalization scale $\mu _R$. This scale at which the mixing angles
and their CTs are given has to be specified explicitly by the user in
the input file of {\texttt{ewN2HDECAY}}. The one-loop corrected
partial decay widths that contain these $\overline{\text{MS}}$ CTs
also depend on the scale $\mu _R$. Moreover, the one-loop partial decay widths
depend on another scale $\mu _\text{out}$ at which the loop integrals
are evaluated. The latter scale should be chosen appropriately to
avoid the appearance of large logarithms in the partial decay
widths. In case that the two scales $\mu _R$ and $\mu _\text{out}$ are
chosen to be different, the parameter conversion automatically
converts the scalar mixing angles from the scale $\mu _R$ to $\mu
_\text{out}$, as further described in
Sec.\,\ref{sec:ParameterConversion}. For this conversion, the
UV-divergent terms of the mixing angle CTs are required, which were
extracted and implemented analytically by isolating the UV-divergent
pieces of the mixing angle CTs as defined in an arbitrary other
renormalization scheme. \\  

\textbf{Adapted KOSY scheme.} 
For the renormalization of the scalar mixing angles in the 2HDM, the
KOSY scheme (denoted by the authors' initials) was proposed in
Ref.\,\cite{Kanemura:2004mg}. It can be directly adapted to the
renormalization of the four mixing angles in the N2HDM as detailed in
Ref.\,\cite{Krause:2017mal}. The adapted KOSY scheme connects the
definition of the scalar mixing angle CTs to the off-diagonal scalar
WFRCs by temporarily switching from the mass basis to the interaction
basis before promoting the mixing angles to one-loop order. 
While the KOSY scheme not only leads to gauge-dependent mixing
angle CTs but also to residual gauge dependences in the partial decay
widths, we implement this scheme for comparative studies with the other
schemes only. Due to the residual gauge dependences, we do not recommend it
for actual use in phenomenological analyses, however. The CTs in the adapted KOSY 
scheme are given by
\begin{mdframed}[frametitle={Renormalization of $\delta \alpha _i$ and
    $\delta \beta$: adapted KOSY scheme (standard
    scheme)},frametitlerule=true,frametitlebackgroundcolor=black!14,frametitlerulewidth=0.6pt,nobreak=true]\begin{align}
\delta \alpha _1 &= \frac{c_{\alpha _3} \left( \text{Re} \left[ \Sigma _{H_1H_2} ( m_{H_1}^2 ) + \Sigma _{H_1H_2} ( m_{H_2}^2 ) \right] - 2\delta T_{H_1H_2} \right) }{2c_{\alpha _2} ( m_{H_1}^2 - m_{H_2}^2 )} \\
&\hspace*{0.45cm} - \frac{s_{\alpha _3} \left( \text{Re} \left[ \Sigma _{H_1H_3} ( m_{H_1}^2 ) + \Sigma _{H_1H_3} ( m_{H_3}^2 ) \right] - 2\delta T_{H_1H_3} \right) }{2c_{\alpha _2} ( m_{H_1}^2 - m_{H_3}^2 )} \nonumber \\
\delta \alpha _2 &= \frac{s_{\alpha _3} \left( \text{Re} \left[ \Sigma _{H_1H_2} ( m_{H_1}^2 ) + \Sigma _{H_1H_2} ( m_{H_2}^2 ) \right] - 2\delta T_{H_1H_2} \right) }{ 2(m_{H_1}^2 - m_{H_2}^2) } \\
&\hspace*{0.45cm} + \frac{c_{\alpha _3} \left( \text{Re} \left[ \Sigma _{H_1H_3} ( m_{H_1}^2 ) + \Sigma _{H_1H_3} ( m_{H_3}^2 ) \right] - 2\delta T_{H_1H_3} \right) }{2( m_{H_1}^2 - m_{H_3}^2) } \nonumber \\
\delta \alpha _3 &= \frac{\text{Re} \left[ \Sigma _{H_2H_3} ( m_{H_2}^2 ) + \Sigma _{H_2H_3} ( m_{H_3}^2 ) \right] - 2\delta T_{H_2H_3} }{ 2(m_{H_2}^2 - m_{H_3}^2) } \\
&\hspace*{0.45cm}- \frac{s_{\alpha _2} c_{\alpha _3} \left( \text{Re} \left[ \Sigma _{H_1H_2} ( m_{H_1}^2 ) + \Sigma _{H_1H_2} ( m_{H_2}^2 ) \right] - 2\delta T_{H_1H_2} \right) }{2c_{\alpha _2} (  m_{H_1}^2 - m_{H_2}^2 ) }  \nonumber \\
&\hspace*{0.45cm}+ \frac{s_{\alpha _2} s_{\alpha _3} \left( \text{Re} \left[ \Sigma _{H_1H_3} ( m_{H_1}^2 ) + \Sigma _{H_1H_3} ( m_{H_3}^2 ) \right] - 2\delta T_{H_1H_3} \right) }{ 2c_{\alpha _2} ( m_{H_1}^2 - m_{H_3}^2 ) } \nonumber \\
\delta \beta ^o &= -\frac{\text{Re} \left[ \Sigma _{G^0A} ( m_A^2 ) + \Sigma _{G^0A} ( 0 ) \right] - 2 \delta T_{G^0A} }{2m_A^2} \\
\delta \beta ^c &= -\frac{\text{Re} \left[ \Sigma _{G^\pm H^\pm} ( m_{H^\pm}^2 ) + \Sigma _{G^\pm H^\pm} (0) \right] - 2\delta T_{G^\pm H^\pm} }{2m_{H^\pm}^2} 
\end{align}\end{mdframed}
\begin{mdframed}[frametitle={Renormalization of $\delta \alpha _i$ and $\delta \beta$: adapted KOSY scheme (alternative FJ scheme)},frametitlerule=true,frametitlebackgroundcolor=black!14,frametitlerulewidth=0.6pt,nobreak=true]\begin{align}
\delta \alpha _1 &= \frac{c_{\alpha _3} \left( \text{Re} \left[ \Sigma ^\text{tad} _{H_1H_2} ( m_{H_1}^2 ) + \Sigma ^\text{tad} _{H_1H_2} ( m_{H_2}^2 ) \right]  \right) }{2c_{\alpha _2} ( m_{H_1}^2 - m_{H_2}^2 )} - \frac{s_{\alpha _3} \left( \text{Re} \left[ \Sigma ^\text{tad} _{H_1H_3} ( m_{H_1}^2 ) + \Sigma ^\text{tad} _{H_1H_3} ( m_{H_3}^2 ) \right]  \right) }{2c_{\alpha _2} ( m_{H_1}^2 - m_{H_3}^2 )} \raisetag{8pt} \\
\delta \alpha _2 &= \frac{s_{\alpha _3} \left( \text{Re} \left[ \Sigma ^\text{tad} _{H_1H_2} ( m_{H_1}^2 ) + \Sigma ^\text{tad} _{H_1H_2} ( m_{H_2}^2 ) \right]  \right) }{ 2(m_{H_1}^2 - m_{H_2}^2) } + \frac{c_{\alpha _3} \left( \text{Re} \left[ \Sigma ^\text{tad} _{H_1H_3} ( m_{H_1}^2 ) + \Sigma ^\text{tad} _{H_1H_3} ( m_{H_3}^2 ) \right]  \right) }{2( m_{H_1}^2 - m_{H_3}^2) } \raisetag{8pt} \\
\delta \alpha _3 &= \frac{\text{Re} \left[ \Sigma ^\text{tad} _{H_2H_3} ( m_{H_2}^2 ) + \Sigma ^\text{tad} _{H_2H_3} ( m_{H_3}^2 ) \right] }{ 2(m_{H_2}^2 - m_{H_3}^2) } - \frac{s_{\alpha _2} c_{\alpha _3} \left( \text{Re} \left[ \Sigma ^\text{tad} _{H_1H_2} ( m_{H_1}^2 ) + \Sigma ^\text{tad} _{H_1H_2} ( m_{H_2}^2 ) \right]  \right) }{2c_{\alpha _2} (  m_{H_1}^2 - m_{H_2}^2 ) } \nonumber \\
&\hspace*{0.45cm}+ \frac{s_{\alpha _2} s_{\alpha _3} \left( \text{Re} \left[ \Sigma ^\text{tad} _{H_1H_3} ( m_{H_1}^2 ) + \Sigma ^\text{tad} _{H_1H_3} ( m_{H_3}^2 ) \right]  \right) }{ 2c_{\alpha _2} ( m_{H_1}^2 - m_{H_3}^2 ) } \raisetag{15pt} \\
\delta \beta ^o &= -\frac{\text{Re} \left[ \Sigma ^\text{tad} _{G^0A} ( m_A^2 ) + \Sigma ^\text{tad} _{G^0A} ( 0 ) \right] }{2m_A^2}  \\
\delta \beta ^c &= -\frac{\text{Re} \left[ \Sigma ^\text{tad} _{G^\pm H^\pm} ( m_{H^\pm}^2 ) + \Sigma ^\text{tad} _{G^\pm H^\pm} (0) \right]  }{2m_{H^\pm}^2} 
\end{align}\end{mdframed}
Like in {\texttt{2HDECAY}},
{\it cf.}\,\cite{Krause:2018wmo}, we implemented two versions of
the KOSY scheme in {\texttt{ewN2HDECAY}} that differ with respect to
the WFRCs through which the CT $\delta \beta$ is defined, namely
$\delta \beta ^o$ and $\delta \beta ^c$ which define the CT through
the WFRCs of the CP-odd and the charged scalar sector, respectively.\\ 

\textbf{$p_{*}$-pinched scheme.} The (adapted) KOSY scheme can be
modified such that a gauge-parameter-independent definition of the
mixing angle CTs is achieved. This approach was suggested in
Refs.\,\cite{Krause:2016gkg, Krause:2016oke} for the 2HDM and in
Ref.\,\cite{Krause:2017mal} for the N2HDM. The derivation of the
mixing angle CTs is analogous to the (adapted) KOSY scheme, but instead
of defining them over the usual off-diagonal WFRCs, the self-energies
are replaced by the pinched self-energies which are derived by means
of the pinch technique (PT), {\it cf.}\,Refs.\,\cite{Binosi:2004qe,
  Binosi:2009qm, Cornwall:1989gv, Papavassiliou:1989zd,
  Degrassi:1992ue, Papavassiliou:1994pr, Watson:1994tn,
  Papavassiliou:1995fq}. For consistency and the cancellation of all
gauge dependences, the alternative tadpole scheme is necessarily
required for this renormalization scheme of the mixing angle CTs. The
pinched scalar self-energies are equivalent to the self-energies
$\Sigma ^\text{tad} (p^2)$ in the alternative tadpole scheme up to
additional self-energy-like contributions $\Sigma ^\text{add}
(p^2)$. In the $p_{*}$-pinched scheme, adapted from
Ref.\,\cite{Espinosa:2002cd} in the Minimal Supersymmetric Extension
of the SM (MSSM), the pinched self-energies
are evaluated at the scale 
\begin{equation}
	p_{*,ij}^2 \equiv \frac{m_i^2 + m_j^2}{2} ~.
\end{equation}
At this scale, the additional self-energy-like contributions $\Sigma
^\text{add}$ vanish. The scalar mixing angle CTs are then given as
follows: 
\begin{mdframed}[frametitle={Renormalization of $\delta \alpha _i$ and $\delta \beta$: $p_{*}$-pinched scheme (alternative FJ scheme)},frametitlerule=true,frametitlebackgroundcolor=black!14,frametitlerulewidth=0.6pt,nobreak=true]\begin{align}
\delta \alpha _1 &= \frac{c_{\alpha _3} \text{Re} \left[ \Sigma ^\text{tad} _{H_1H_2} (p_{*,12}^2 ) \right] _{\xi = 1} }{c_{\alpha _2} ( m_{H_1}^2 - m_{H_2}^2 )} - \frac{s_{\alpha _3} \text{Re} \left[ \Sigma ^\text{tad} _{H_1H_3} (p_{*,13}^2 ) \right] _{\xi = 1}}{c_{\alpha _2} ( m_{H_1}^2 - m_{H_3}^2 )}  \\
\delta \alpha _2 &= \frac{s_{\alpha _3} \text{Re} \left[ \Sigma ^\text{tad} _{H_1H_2} (p_{*,12}^2 ) \right] _{\xi = 1}}{ m_{H_1}^2 - m_{H_2}^2 } + \frac{c_{\alpha _3} \text{Re} \left[ \Sigma ^\text{tad} _{H_1H_3} (p_{*,13}^2 ) \right] _{\xi = 1} }{ m_{H_1}^2 - m_{H_3}^2 } \\
\delta \alpha _3 &= \frac{\text{Re} \left[ \Sigma ^\text{tad} _{H_2H_3} (p_{*,23}^2 ) \right] _{\xi = 1}}{ m_{H_2}^2 - m_{H_3}^2 } + \frac{s_{\alpha _2} s_{\alpha _3} \text{Re} \left[ \Sigma ^\text{tad} _{H_1H_3} (p_{*,13}^2 ) \right] _{\xi = 1} }{ c_{\alpha _2} ( m_{H_1}^2 - m_{H_3}^2 ) } \\
&\hspace*{0.45cm}- \frac{s_{\alpha _2} c_{\alpha _3} \text{Re} \left[ \Sigma ^\text{tad} _{H_1H_2} (p_{*,12}^2 ) \right] _{\xi = 1} }{c_{\alpha _2} (  m_{H_1}^2 - m_{H_2}^2 ) }  \nonumber \\
\delta \beta ^o &= -\frac{1}{m_A^2} \text{Re} \left[ \Sigma ^\text{tad} _{G^0A} \left( \frac{m_A^2 }{2} \right) \right] _{\xi = 1} \\
\delta \beta ^c &= -\frac{1}{m_{H^\pm}^2} \text{Re} \left[ \Sigma ^\text{tad} _{G^\pm H^\pm} \left( \frac{m_{H^\pm}^2 }{2} \right) \right] _{\xi = 1} 
\end{align}\end{mdframed}
As for the adapted KOSY scheme, we implemented two different
variations of the $p_{*}$-pinched scheme that differ in the definition
of the CT of $\beta$. The index '$\xi = 1$' means that the self-energies
are evaluated in the Feynman gauge. \\ 

\textbf{OS-pinched scheme.} In the OS-pinched scheme, the pinched
scalar self-energies are evaluated at the OS-inspired mass scale of the
corresponding scalar particle. In this case, the
additional UV-divergent\footnote{While the additional self-energy-like
  contributions $\Sigma ^\text{add}$ are all separately UV-divergent,
  they appear only in UV-finite combinations in the definition of the
  mixing angle CTs $\delta \alpha _i$ ($i=1,2,3$),
  {\it cf.}\,Ref.\,\cite{Krause:2018wmo}.} self-energy-like contributions
$\Sigma ^\text{add}$ do not vanish. They were derived for the N2HDM in
Ref.\,\cite{Krause:2018wmo} and read 
\begin{align}
\Sigma ^\textrm{add} _{H_iH_j} (p^2) &= -\frac{\alpha _\text{em} m_Z^2}{8 \pi m_W^2 \left( 1 - \frac{m_W^2}{m_Z^2} \right) } \left( p^2 - \frac{ m_{H_i}^2 + m_{H_j}^2}{2} \right) \label{eq:RenormalizationScalarAnglesAdditionalTermCPEven} \\
&\hspace*{0.4cm} \cdot \bigg\{ \mathcal{O}^{(1)}_{H_iH_j} B_0( p^2; m_Z^2, m_{A}^2) + \mathcal{O}^{(2)}_{H_iH_j} B_0( p^2; m_Z^2, m_{Z}^2)  \nonumber \\
&\hspace*{0.8cm} + 2\frac{m_W^2}{m_Z^2} \left[ \mathcal{O}^{(1)}_{H_iH_j} B_0( p^2; m_W^2, m_{H^\pm }^2) + \mathcal{O}^{(2)}_{H_iH_j} B_0( p^2; m_W^2, m_{W}^2) \right] \bigg\}  \nonumber \\
\Sigma ^\textrm{add} _{G^0 A } (p^2) &= -\frac{\alpha _\text{em} m_Z^2}{8 \pi m_W^2 \left( 1 - \frac{m_W^2}{m_Z^2} \right) } \left( p^2 - \frac{m_{A}^2}{2} \right) \sum _{k=1}^3 \mathcal{O}^{(3)}_{H_kH_k} B_0( p^2; m_Z^2, m_{H_k}^2) \label{eq:RenormalizationScalarAnglesAdditionalTermCPOdd} \\
\Sigma ^\textrm{add} _{G^\pm H^\pm } (p^2) &= -\frac{\alpha _\text{em} }{4 \pi \left( 1 - \frac{m_W^2}{m_Z^2} \right) } \left( p^2 - \frac{m_{H^\pm }^2}{2} \right) \sum _{k=1}^3 \mathcal{O}^{(3)}_{H_kH_k} B_0( p^2; m_Z^2, m_{H_k}^2)  \;,\label{eq:RenormalizationScalarAnglesAdditionalTermCharged} 
\end{align}
where $i,j=1,2,3$ and $\mathcal{O}^{(x)}_{H_iH_j}$ ($x=1,...,4$) is a
shorthand notation for the following combinations of coupling
constants between the Higgs and gauge sectors as defined in
Tab.\,\ref{tab:higgsGaugeCouplingDefinitions}: 
\begin{align}
\mathcal{O}^{(1)}_{H_iH_j} &= \tilde{\kappa } _{H_iVH} \cdot \tilde{\kappa } _{H_jVH} \\
\mathcal{O}^{(2)}_{H_iH_j} &= \kappa _{H_iVV} \cdot \kappa _{H_jVV} \\
\mathcal{O}^{(3)}_{H_iH_j} &= \kappa _{H_iVV} \cdot \tilde{\kappa } _{H_jVH} ~.
\end{align}
The mixing angle CTs in the OS-pinched scheme are given by:
\begin{mdframed}[frametitle={Renormalization of $\delta \alpha _i$ and $\delta \beta$: OS-pinched scheme (alternative FJ scheme)},frametitlerule=true,frametitlebackgroundcolor=black!14,frametitlerulewidth=0.6pt,nobreak=true]\begin{align}
\delta \alpha _1 &= \frac{c_{\alpha _3} \left( \text{Re} \left[ \Sigma ^\text{tad} _{H_1H_2} ( m_{H_1}^2 ) + \Sigma ^\text{tad} _{H_1H_2} ( m_{H_2}^2 ) \right] _{\xi = 1} + \Sigma ^\text{add} _{H_1H_2} ( m_{H_1}^2 ) + \Sigma ^\text{add} _{H_1H_2} ( m_{H_2}^2 ) \right) }{2c_{\alpha _2} ( m_{H_1}^2 - m_{H_2}^2 )} \\
&\hspace*{0.45cm} - \frac{s_{\alpha _3} \left( \text{Re} \left[ \Sigma ^\text{tad} _{H_1H_3} ( m_{H_1}^2 ) + \Sigma ^\text{tad} _{H_1H_3} ( m_{H_3}^2 ) \right] _{\xi = 1} + \Sigma ^\text{add} _{H_1H_3} ( m_{H_1}^2 ) + \Sigma ^\text{add} _{H_1H_3} ( m_{H_3}^2 ) \right) }{2c_{\alpha _2} ( m_{H_1}^2 - m_{H_3}^2 )} \nonumber \\
\delta \alpha _2 &= \frac{s_{\alpha _3} \left( \text{Re} \left[ \Sigma ^\text{tad} _{H_1H_2} ( m_{H_1}^2 ) + \Sigma ^\text{tad} _{H_1H_2} ( m_{H_2}^2 ) \right] _{\xi = 1} + \Sigma ^\text{add} _{H_1H_2} ( m_{H_1}^2 ) + \Sigma ^\text{add} _{H_1H_2} ( m_{H_2}^2 ) \right) }{ 2(m_{H_1}^2 - m_{H_2}^2) } \\
&\hspace*{0.45cm} + \frac{c_{\alpha _3} \left( \text{Re} \left[ \Sigma ^\text{tad} _{H_1H_3} ( m_{H_1}^2 ) + \Sigma ^\text{tad} _{H_1H_3} ( m_{H_3}^2 ) \right] _{\xi = 1} + \Sigma ^\text{add} _{H_1H_3} ( m_{H_1}^2 ) + \Sigma ^\text{add} _{H_1H_3} ( m_{H_3}^2 ) \right)  }{2( m_{H_1}^2 - m_{H_3}^2) } \nonumber \\
\delta \alpha _3 &= \frac{\text{Re} \left[ \Sigma ^\text{tad} _{H_2H_3} ( m_{H_2}^2 ) + \Sigma ^\text{tad} _{H_2H_3} ( m_{H_3}^2 ) \right] _{\xi = 1} + \Sigma ^\text{add} _{H_2H_3} ( m_{H_2}^2 ) + \Sigma ^\text{add} _{H_2H_3} ( m_{H_3}^2 )  }{ 2(m_{H_2}^2 - m_{H_3}^2) } \\
&\hspace*{0.45cm}- \frac{s_{\alpha _2} c_{\alpha _3} \left( \text{Re} \left[ \Sigma ^\text{tad} _{H_1H_2} ( m_{H_1}^2 ) + \Sigma ^\text{tad} _{H_1H_2} ( m_{H_2}^2 ) \right] _{\xi = 1} + \Sigma ^\text{add} _{H_1H_2} ( m_{H_1}^2 ) + \Sigma ^\text{add} _{H_1H_2} ( m_{H_2}^2 ) \right) }{2c_{\alpha _2} (  m_{H_1}^2 - m_{H_2}^2 ) }  \nonumber \\
&\hspace*{0.45cm}+ \frac{s_{\alpha _2} s_{\alpha _3} \left( \text{Re} \left[ \Sigma ^\text{tad} _{H_1H_3} ( m_{H_1}^2 ) + \Sigma ^\text{tad} _{H_1H_3} ( m_{H_3}^2 ) \right] _{\xi = 1} + \Sigma ^\text{add} _{H_1H_3} ( m_{H_1}^2 ) + \Sigma ^\text{add} _{H_1H_3} ( m_{H_3}^2 ) \right) }{ 2c_{\alpha _2} ( m_{H_1}^2 - m_{H_3}^2 ) } \nonumber \\
\delta \beta ^o &= -\frac{\text{Re} \left[ \Sigma ^\text{tad} _{G^0A} ( m_A^2 ) + \Sigma ^\text{tad} _{G^0A} ( 0 ) \right] _{\xi = 1} + \Sigma ^\text{add} _{G^0A} ( m_A^2 ) + \Sigma ^\text{add} _{G^0A} ( 0 )}{2m_A^2} \\
\delta \beta ^c &= -\frac{\text{Re} \left[ \Sigma ^\text{tad} _{G^\pm H^\pm} ( m_{H^\pm}^2 ) + \Sigma ^\text{tad} _{G^\pm H^\pm} (0) \right] _{\xi = 1} + \Sigma ^\text{add} _{G^\pm H^\pm} ( m_{H^\pm}^2 ) + \Sigma ^\text{add} _{G^\pm H^\pm} ( 0 )}{2m_{H^\pm}^2} 
\end{align}\end{mdframed}
\begin{table}[tb]
\centering
  \begin{tabular}{ c c c }
    \hline
    & $\kappa _{H_iVV}$ & $\tilde{\kappa} _{H_iVH}$ \\ \hline
    $H_1$ & $c_{\alpha _2} c_{\beta - \alpha _1}$ & $-c_{\alpha _2} s_{\beta - \alpha _1}$ \\
    $H_2$ & $-s_{\alpha _2} s_{\alpha _3} c_{\beta - \alpha _1} + c_{\alpha _3} s_{\beta - \alpha _1}$ & $s_{\alpha _2} s_{\alpha _3} s_{\beta - \alpha _1} + c_{\alpha _3} c_{\beta - \alpha _1}$ \\
    $H_3$ & $-s_{\alpha _2} c_{\alpha _3} c_{\beta - \alpha _1} - s_{\alpha _3} s_{\beta - \alpha _1}$ & $s_{\alpha _2} c_{\alpha _3} s_{\beta - \alpha _1} - s_{\alpha _3} c_{\beta - \alpha _1}$ \\
    \hline
  \end{tabular}
    \caption{Coupling constants for the Higgs-gauge sector, as defined
      in Ref.\,\cite{Krause:2017mal}, where $\kappa _{H_iVV}$ denotes
      the coupling of a CP-even Higgs boson $H_i$ ($i=1,2,3$) to a
      pair of gauge bosons $V=W^\pm,Z$ and $\tilde{\kappa} _{H_iVH}$
      denotes the coupling of a CP-even Higgs boson to a gauge boson
      and an additional Higgs boson $H=A,H^\pm$.} 
   \label{tab:higgsGaugeCouplingDefinitions}
\end{table}

\subsubsection{Renormalization of the Fermion Sector}
\label{sec:renormalizationFermionSector}
The renormalization of the fermion sector in the N2HDM is performed as
in the 2HDM. All CTs of the fermion masses, the 
CKM mixing matrix elements and the fermion WFRCs are implemented with
definitions analogous to those presented in \cite{Krause:2018wmo}. The only
difference with respect to the 2HDM are the different Yukawa couplings
due to the extended scalar sector of the N2HDM. The CTs of the Yukawa
coupling parameters defined in Tab.\,\ref{tab:yukawaCouplings} are
given by the following relations for $f=l,u,d$ which hold independently
of the chosen N2HDM type: 
\begin{align}
\delta Y^f_1 &= c_{\alpha _2} \left( c_{\alpha _3} Y^f_2 - s_{\alpha _3} Y^f_3 \right) \delta \alpha _1 - t_{\alpha _2} Y^f_1 \delta \alpha _2 - Y^f_1 Y^f_4 \delta \beta \\
\delta Y^f_2 &= \left( s_{\alpha _2} Y^f_3 - c_{\alpha _2} c_{\alpha _3} Y^f_1 \right) \delta \alpha _1 - s_{\alpha _3} Y^f_1 \delta \alpha _2 + Y^f_3 \delta \alpha _3 - Y^f_2 Y^f_4 \delta \beta \\
\delta Y^f_3 &= \left( c_{\alpha _2} s_{\alpha _3} Y^f_1 - s_{\alpha _2} Y^f_2 \right) \delta \alpha _1 - c_{\alpha _3} Y^f_1 \delta \alpha _2 - Y^f_2 \delta \alpha _3 - Y^f_3 Y^f_4 \delta \beta \\
\delta Y^f_4 &= - \left( 1 + \left(Y^f_4 \right)^2 \right) \delta \beta ~.
\end{align}

\subsubsection{Renormalization of the Soft-$\mathbb{Z}_2$-Breaking Parameter $m_{12}^2$} 
\label{sec:renormalizationSoftm12Squared}
The soft-$\mathbb{Z}_2$-breaking Parameter $m_{12}^2$ is promoted to NLO by introducing a CT $\delta m_{12}^2$ according to
\begin{equation}
m_{12}^2 ~\rightarrow ~ m_{12}^2 + \delta m_{12}^2 ~.
\end{equation}
One possibility to fix the CT for $m_{12}^2$ is to define it via a
Higgs-to-Higgs decay, since the Higgs self-couplings contain the
parameter at tree level. However, such a process-dependent definition
of $\delta m_{12}^2$ has the drawback that the CT is defined not only
via the genuine vertex corrections of the Higgs-to-Higgs decay, but
additionally as a function of several other CTs of the N2HDM due to
the intricate structure of the Higgs self-couplings. As a result, such
a CT can introduce very large finite contributions which leads to NLO
corrections that can become very large as well. This has already been observed
in Ref.\,\cite{Krause:2016xku} in the context of the 2HDM. We
therefore do not implement a process-dependent scheme for $\delta
m_{12}^2$ in {\texttt{ewN2HDECAY}} but instead fix the CT through an
$\overline{\mbox{MS}}$ condition, \textit{i.e.}\,$\delta m_{12}^2$
only contains UV-divergent and some global finite parts parts
proportional to 
\begin{equation}
	\Delta \equiv \frac{1}{\varepsilon } - \gamma _E + \ln (4\pi )
        + \ln \left( \frac{\mu ^2}{\mu _R ^2} \right) \;,
\end{equation}
where $\gamma _E$ denotes the Euler-Mascheroni constant, $\mu$ denotes
the mass-dimensional 't Hooft scale which cancels in the calculation
of decay amplitudes and the regulator $\varepsilon$ is introduced in
the framework of dimensional regularization, {\it
  cf.}\,Refs.\,\cite{Wilson:1970ag,Wilson:1971dc,Ashmore1972,Bollini:1972ui,THOOFT1972189}. 
In {\texttt{ewN2HDECAY}}, we extracted the UV divergence of $\delta
m_{12}^2$ by calculating the one-loop amplitude of the decay $H_1
\rightarrow AA$, including the genuine vertex corrections of the
process as well as all CTs apart from $\delta m_{12}^2$, and by
considering the residual UV divergence which is then assigned to
$\delta m_{12}^2$. This yields the following analytic expression of
the CT,
\begin{mdframed}[frametitle={Renormalization of $m_{12}^2$ (standard
    and alternative FJ scheme)},frametitlerule=true,frametitlebackgroundcolor=black!14,frametitlerulewidth=0.6pt,nobreak=true]
\begin{align}
\delta m_{12}^2 &= \frac{\alpha _\text{em} m_{12}^2 }{16\pi m_W^2 \left( 1 - \frac{m_W^2}{m_Z^2} \right)} \Big[ \frac{8m_{12}^2}{s_{2\beta }} - 2m_{H^\pm }^2 - m_A^2 + \sum _{i=1}^3 R_{i1}R_{i2}m_{H_i}^2 - 3(2m_W^2 + m_Z^2) \label{eq:renormalizationConditionm12Sq} \raisetag{1.8\baselineskip} \\
&\hspace*{0.5cm} + \sum _u 6 m_u^2  Y^u_4 \left( Y^u_4 - \frac{1}{t_{2\beta}} \right)+ \sum _d 6 m_d^2  Y^d_4 \left( Y^d_4 - \frac{1}{t_{2\beta}} \right) + \sum _l 2 m_l^2  Y^l_4 \left( Y^l_4 - \frac{1}{t_{2\beta}} \right) \Big] \Delta  \nonumber
\end{align}\end{mdframed}
where the sums are performed over all up- and down-type quarks as well
as over all charged leptons, respectively. Since $m_{12}^2$ is a
genuine parameter of the N2HDM Higgs potential before EWSB, it is
gauge-independent and the CT is invariant under a change of the
tadpole renormalization so that $\delta m_{12}^2$ is the same in both
tadpole schemes. Due to the $\overline{\text{MS}}$ renormalization of
$\delta m_{12}^2$, the CT explicitly depends on the renormalization
scale $\mu _R$ whose value has to be specified by the user\footnote{In
  {\texttt{ewN2HDECAY}}, all $\overline{\text{MS}}$ parameters are
  given at the same global scale $\mu _R$.}. In case that the
renormalization scale $\mu _\text{out}$ at which the one-loop partial
decay widths are evaluated differs from the input renormalization
scale $\mu _R$, the parameter conversion routine,
{\it cf.}\,Sec.\,\ref{sec:ParameterConversion}, evolves the parameter
$m_{12}^2$ from $\mu _R$ to $\mu _\text{out}$, analogous to the
$\overline{\text{MS}}$ renormalized scalar mixing angles. 

\subsubsection{Renormalization of the singlet VEV $v_S$} 
While the singlet VEV $v_S$ is already accounted for by the proper
renormalization of the minimum of the Higgs potential, {\it cf.}\,Sec.\,\ref{sec:renormalizationTadpoles}, it still receives an additional CT after the minimum conditions are imposed on the one-loop potential. This is in analogy to the doublet VEVs. In the framework of \textit{e.g.}\,the alternative tadpole scheme, the doublet VEVs $v_1$ and $v_2$ acquire shifts according to \eqref{eq:vevShifts}. These shifts ensure that the VEVs are equivalent to the gauge-independent tree-level VEVs of the potential, so that
\begin{equation}
	\left. v^\text{ren} \right| _\text{FJ} = v^\text{tree} = \left. \frac{2m_W}{g} \right| ^\text{tree}
\end{equation}
holds. After the minimum conditions are applied through the VEV
shifts, the tree-level parameters $m_W$ and $g$ still need to be
renormalized, so that $v$ effectively acquires an additional CT
$\Delta v$ in form of a combination of CTs for $m_W^2$ and $g$,
\begin{equation}
\left. \frac{2m_W}{g} \right| ^\text{tree} \rightarrow
\left. \frac{2m_W}{g} \right| ^\text{ren}_\text{FJ} + \underbrace{
  \left. \frac{2m_W}{g} \left( \frac{\delta m_W^2}{2m_W^2} -
      \frac{\delta g}{g} \right) \right| _\text{FJ} }_{\equiv \Delta v} ~.
\end{equation}
Analogously, the singlet VEV $v_S$ is shifted according to
\eqref{eq:vevShifts}, ensuring that $v_S$ is equivalent to the
tree-level singlet VEV in the alternative tadpole scheme, and after
the minimum conditions of the potential are applied at one-loop order,
$v_S$ acquires an additional CT $\Delta v_S$ through 
\begin{equation}
	\left. v_S \right| ^\text{tree} \rightarrow \left. v_S \right| ^\text{ren}_\text{FJ} + \Delta v_S ~.
\end{equation}
For more details about the appearance of these additional CTs, we
refer to Ref.\,\cite{Krause:2017mal}.  Transferring the general
analysis on the renormalization of spontaneously broken gauge
symmetries presented in Ref.\,\cite{Sperling:2013eva} in the framework
of the standard tadpole scheme to the N2HDM 
yields the conclusion that $\Delta v_S$ cannot contain UV divergences
since the corresponding singlet field obeys a rigid
invariance. Therefore, $\Delta v_S$ contains at most finite parts in
the standard tadpole scheme 
which can be fixed \textit{e.g.}\,in a process-dependent
renormalization scheme. In {\texttt{ewN2HDECAY}}, we implemented
$\overline{\text{MS}}$ conditions for the singlet VEV CT in order to
avoid potentially large finite contributions that would be introduced
in the CT when fixing it through a Higgs-to-Higgs decay. The CT is
implemented by setting its finite part $\left. \Delta v_S
\right|_\text{fin}$ to zero: 
\begin{mdframed}[frametitle={Renormalization of the tree-level $v_S$ (standard and alternative FJ scheme)},frametitlerule=true,frametitlebackgroundcolor=black!14,frametitlerulewidth=0.6pt,nobreak=true]
\begin{align}
\left. \Delta v_S \right| _\text{fin} = 0
\end{align}\end{mdframed}
As for $\overline{\text{MS}}$ renormalized scalar mixing angles and
$m_{12}^2$, the value of $v_S$ is converted from the input
renormalization scale $\mu _R$ to the scale $\mu _\text{out}$ at which
the decays are evaluated in case that both scales differ. For this
conversion the UV-divergent parts of $\Delta v_S$ are needed. In the
standard tadpole scheme, these UV-divergent parts are identically zero
due to the rigid invariance, and as a consequence, the parameter $v_S$
is not converted and remains the same if the two scales are
different\footnote{This corresponds to the vanishing of the one-loop
  beta function of $v_S$ in the standard tadpole scheme.}. In the
alternative tadpole scheme, however, $\Delta v_S$ contains
additional UV divergences which have been extracted by calculating the
remaining UV-divergent parts of the off-shell process $H_1 \rightarrow
H_1 H_1$ to one-loop order when all CTs apart from $\Delta v_S$ are
fixed. Due to the lengthy analytic expression of the UV-divergent
parts of $\Delta v_S$, we do not state them explicitly here.  

\subsection{Electroweak Decay Processes at LO and NLO}
\label{sec:decayProcessesAtLOandNLO}
For the EW corrections, we only consider OS decay processes, {\it
  i.e.}~the decays of all Higgs bosons with four-momentum $p_1$ into
two particles $X_1$ and $X_2$ with four-momenta $p_2$ and $p_3$,
respectively, for which
\begin{equation}
	p_1^2 \geq \left( p_2 + p_3 \right) ^2 ~.
\end{equation}
Note that we do not include here loop-induced tree-level decays, so
that the EW one-loop corrections are calculated for the following
processes ($i=1,2,3$),
\begin{itemize}
	\item $H_i/A \to f\bar{f}$ ~ ($f=c,s,t,b,\mu ,\tau $) 
	\item $H_i \to VV$ ~ ($V=W^\pm ,Z$) 
	\item $H_i \to VS$ ~ ($VS=ZA, W^\pm H^\mp$) 
	\item $H_i \to SS$ ~ ($i=1,2,3$, $S = A, H^\pm$) 
	\item $H_i \to H_jH_k$ ~ ($i,j,k=1,2,3$ and $i>j,k$)
	\item $A \to VS$ ~ ($VS=Z H_i, W^\pm H^\mp$) 
	\item $H^+ \to f\bar{f}'$ ~ ($f\bar{f}'=u/c/t + \bar{b}, u/c/t
          + \bar{s}, c/t + \bar{d}, \nu_\tau \bar{\tau}, \nu_\mu \bar{\mu}$)  
	\item $H^\pm \to VS$ ~ ($VS=W^\pm H_i, W^\pm A$) 
\end{itemize}
Like in {\texttt{2HDECAY}}, we do not consider decays containing
first-generation fermions, \textit{i.e.}\,the
decays $H_i/A \to f\bar{f}$ ($i=1,2,3$, $f=u,d,e $) and $H^+ \to
f\bar{f}$ ($f\bar{f}=u\bar{d}, \nu _e e^+  $) are neglected in
{\texttt{ewN2HDECAY}}. 

From a technical point of view, the calculation of the LO and NLO
decay amplitudes in {\texttt{ewN2HDECAY}} is identical to their
calculation in the 2HDM in {\texttt{2HDECAY}}. In the following, we
only briefly mention the tools used for the calculation. For a more
detailed explanation of the calculation, we refer to
Ref.\,\cite{Krause:2018wmo}. The Feynman amplitudes for the LO and NLO
decays as well as for the tadpole diagrams and self-energies required
for the definition of the CTs were generated with {\texttt{FeynArts
    3.9}} \cite{Hahn:2000kx}. The N2HDM model file necessary for the
generation of the amplitudes was created with {\texttt{SARAH 4.14.0}}
\cite{Staub:2010jh,Staub:2012pb,Staub:2013tta,Goodsell:2014bna,Goodsell:2014pla}. The 
simplification of the Dirac algebra, the Passarino-Veltman reduction
and the analytic evaluation of the decay amplitudes as well as the CTs
was performed with the help of the tool {\texttt{FeynCalc 8.2.0}}
\cite{MERTIG1991345, Shtabovenko:2016sxi}. The real corrections
necessary for the cancellation of all infra-red (IR) divergences were
implemented analytically by applying the general results given in
Ref.\,\cite{Goodsell:2017pdq} to the N2HDM. For the evaluation of the
integrals involved in the real corrections, the analytic expressions
of Ref.\,\cite{Denner:1991kt} were implemented. The
numerical evaluation of the loop integrals is performed by linking
{\texttt{LoopTools 2.14}} \cite{HAHN1999153}. 

\subsection{Link to N2HDECAY \label{sec:n2hdecaylink}}
\label{sec:connectionHDECAY}
For our new tool {\texttt{ewN2HDECAY}} we combine the EW one-loop
corrections to the N2HDM Higgs decays with 
the Fortran code {\texttt{N2HDECAY}}. This code is based on an
extension of the Fortran code {\texttt{HDECAY}} version 6.511 \cite{DJOUADI199856,
  Djouadi:2018xqq} to the N2HDM
\cite{Muhlleitner:2016mzt,Engeln:2018mbg}, which includes the
state-of-the-art QCD corrections in the partial decay
widths.\footnote{Details can be found in
  \cite{DJOUADI199856,Djouadi:2018xqq}.} Care has to be taken when
combining the two codes in order to remain consistent at higher loop
level. We commented on this in detail in \cite{Krause:2018wmo} and
summarise here only very briefly the main points. 

In order to consistently combine {\tt (N2)HDECAY} which uses $\{ G_F, m_W,
m_Z \}$ as independent input parameters with the EW-corrected decays
based on the set $\{ \alpha_{\text{em}} (m_Z), m_W, m_Z\}$ as independent
input parameters, we choose a pragmatic solution where the {\tt
  N2HDECAY} decay widths in terms of $G_F$ are rescaled with
$G_F^\text{calc}/G_F$, with $G_F^\text{calc}$  
being calculated through the tree-level relation
\eqref{eq:definitionFermiConstant} as a function of $\alpha_\text{em}
(m_Z^2)$.\footnote{The proper conversion between the $\{ G_F, m_W,
m_Z \}$ and $\{ \alpha_{\text{em}} (m_Z), m_W, m_Z\}$ scheme would
require the inclusion of N2HDM higher-order corrections in the
conversion formulae. We expect the differences with respect to our
pragmatic approach to be small.} In case the user does not choose to
calculate the EW corrections no such rescaling is performed. Note also
that {\tt N2HDECAY} includes off-shell decays in certain final
states. In {\tt ewN2HDECAY} the EW and QCD corrections are combined
such that {\tt N2HDECAY} computes the off-shell decay widths, but the
EW corrections are only added to OS decays. We furthermore assume that
the QCD and EW corrections factorize. The relative QCD corrections
$\delta^{\text{QCD}}$ are defined with respect to the LO width
$\Gamma^\text{N2HD,LO}$ that is calculated by {\tt N2HDECAY} and contains
{\it e.g.}~also running quark masses in order to improve the
perturbative behaviour. The relative EW corrections
$\delta^{\text{EW}}$, however, are obtained by normalizing to the LO
width with OS particle masses. 
The QCD and EW corrected decay width into a specific final
state, $\Gamma^{\text{QCD\&EW}}$, is hence calculated as 
\beq
\Gamma^{\text{QCD\&EW}} = \frac{G_F^{\text{calc}}}{G_F} \Gamma^{\text{N2HD,LO}} 
[1+\delta^{\text{QCD}}] [1+ + \delta^{\text{EW}}] 
\equiv \frac{G_F^\text{calc}}{G_F}
\Gamma^{\text{N2HD,QCD}} 
[1 + \delta^{\text{EW}}]  \;. \label{eq:ewqcdwidths}
\eeq
The formula includes the aforementioned rescaling factor $G_F^\text{calc}/G_F$.

For each input file generically called {\texttt{inputfilename.in}} in
the following\footnote{The input filename can be
    chosen arbitrarily by the user.}, the program {\tt ewN2HDECAY}
provides two separate output files which adopt the filename of the
input file together with additional suffices as explained in the
following. The file {\tt inputfilename\_BR.out} contains both the
total widths and branching ratios as calculated by {\tt N2HDECAY} without the EW
corrections\footnote{We remind the reader that they include
  loop-induced and off-shell decays as well as QCD
  corrections where applicable. They are furthermore rescaled by
  $G_F^{\text{calc}}/G_F$ in case the flag for the EW corrections is
  turned on in the input file.} and 
those including the EW corrections based on the formula
Eq.~(\ref{eq:ewqcdwidths}). In the file {\tt inputfilename\_EW.out} the
LO and EW-corrected NLO decay widths are given out. Note that the LO
widths do not contain any running quark masses here in decays into
quark pair final states. Furthermore, the widths are calculated in the 
$\{\alpha_{\text{em}}, m_W, m_Z\}$ scheme. 
The widths given out here are not useful for phenomenological analyses
as they do not include any QCD corrections nor loop-induced or
off-shell decays. They can be used, however, for the study on the
importance of the EW corrections and an estimate of the remaining
theoretical uncertainty due to missing higher-order corrections by
changing the renormalization schemes.

We finally remark that for certain parameter choices the EW-corrected
decay widths can become negative. This can be due to a small LO
width, due to an artificial enhancement of the EW corrections because
of a badly chosen renormalization scheme or due to parametrically
enhanced EW corrections because of large involved couplings. In this
case, the NLO corrections cannot be trusted of course and should be
discarded. For further details and discussions, we refer to
\cite{Krause:2016oke,Krause:2016xku,Krause:2017mal,Krause:2018wmo}.

\subsection{Parameter Conversion}
\label{sec:ParameterConversion}
The one-loop corrected partial decay widths explicitly depend on the
renormalization scale $\mu _\text{out}$ at which the integrals of the
higher-order corrections are evaluated. In {\texttt{ewN2HDECAY}}, this
scale can be chosen by the user to be either at a fixed global value
or to be equal to the mass of the decaying particle. Additionally, all
$\overline{\text{MS}}$ renormalized parameters introduce a dependence
on the input renormalization scale $\mu _R$ at which these parameters
are defined. This scale has to be given by the user in the input
file. In case the two scales $\mu _R$ and $\mu _\text{out}$ differ,
the $\overline{\text{MS}}$ parameters need to be converted from the
input scale $\mu _R$ to the output scale $\mu _\text{out}$, which is done by means of
the formula  
\begin{equation}
	\varphi \left( \{ \mu _\text{out} \} \right) \approx \varphi \left( \{ \mu _R \} \right) + \ln \left( \frac{\mu _\text{out}^2}{\mu _R^2} \right) \delta \varphi ^\text{div} \left( \{ \varphi \} \right) \label{eq:scalechange}
\end{equation}
where $\varphi$ and $\delta \varphi$ denote all $\overline{\text{MS}}$
parameters ($m_{12}^2$, $v_S$ and $\alpha _i$ ($i=1,2,3$) and $\beta$,
if the latter are defined in an $\overline{\text{MS}}$ scheme) and
their respective CTs. The superscript 'div' indicates that only the
UV-divergent part of the CT is taken into account. 

Apart from the conversion of the $\overline{\text{MS}}$ parameters
from one scale to another, an additional parameter conversion of the
scalar mixing angles has to be performed if the renormalization scheme
at which their input is given is different from the renormalization
scheme with which the one-loop partial decay widths are
evaluated. Since the 10 different renormalization schemes implemented
in {\texttt{ewN2HDECAY}} differ only in their definition of the mixing
angle CTs, the scalar mixing angles are the only parameters affected
by the parameter conversion. The values of the mixing angles given in
the reference scheme, \textit{i.e.}\,$\varphi _\text{ref}$, and their
values $\varphi_i$ in another different renormalization scheme are
connected to each other via the bare values of the mixing angles,
which are independent of the renormalization scheme. Together with the
corresponding CTs $\delta \varphi _\text{ref}$ and $\delta \varphi _i$
in the two different schemes, the conversion of the values is given by 
\begin{equation}
	\varphi _i (\{  \mu_{\text{out}}\}) \approx \varphi
        _\text{ref} (\{  \mu_R \}) + \delta \varphi
        _\text{ref} \left( \{ \varphi _\text{ref}, \mu_R \} \right) - \delta
        \varphi _i \left( \{ \varphi _\text{ref},\mu_{\text{out}} \}
        \right) \;. \label{eq:convertedParameterValues} 
\end{equation}
This linearized relation holds approximately up to higher-order terms,
since the CT $\delta \varphi _i$ on the right-hand side is evaluated
with the parameters $\varphi$ given in the reference scheme instead of
the other renormalization scheme in which $\varphi _i$ is defined. We
want to emphasize that \eqref{eq:convertedParameterValues} also
contains a dependence on the scales $\mu _R$ and $\mu _\text{out}$
which is of importance if the scalar mixing angels are defined in an
$\overline{\text{MS}}$ scheme. 

\section{Program Description}
\label{sec:programDescriptionMain}
In this section, we describe the system requirements for
{\texttt{ewN2HDECAY}} and provide a guide for installing and using the
program. Moreover, we describe the format of the input and output
files in detail. 
\subsection{System Requirements}
The \texttt{Python/FORTRAN} program code {\texttt{ewN2HDECAY}} was
developed under {\texttt{Windows 10}} and {\texttt{openSUSE Leap
    15.0}}. The following operating systems are supported:
\begin{itemize}
	\item {\texttt{Windows 7}} and {\texttt{Windows 10}} (tested
          with {\texttt{Cygwin 2.10.0}}) 
	\item {\texttt{Linux}} (tested with {\texttt{openSUSE Leap 15.0}})
	\item {\texttt{macOS}} (tested with {\texttt{macOS Sierra 10.12}})
\end{itemize}
The compilation of {\texttt{ewN2HDECAY}} under {\texttt{Windows}}
requires an installed up-to-date version of {\texttt{Cygwin}}
(together with the packages {\texttt{cURL}}, {\texttt{find}},
{\texttt{gcc}}, {\texttt{g++}} and {\texttt{gfortran}}). For the
compilation of {\texttt{ewN2HDECAY}} the 
{\texttt{GNU C}} compilers {\texttt{gcc}} (tested with versions
{\texttt{6.4.0}} and {\texttt{7.3.1}}), {\texttt{g++}} and the
{\texttt{FORTRAN}} compiler {\texttt{gfortran}} are needed. Additionally, an
up-to-date version of either {\texttt{Python 2}} or {\texttt{Python
    3}} are required (tested with versions {\texttt{2.7.14}} and
{\texttt{3.5.0}}). 
\subsection{License}
{\texttt{ewN2HDECAY}} is released under the GNU General Public License
(GPL) ({\texttt{GNU GPL-3.0-or-later}}). {\texttt{ewN2HDECAY}} is free
software, which means that anyone can redistribute it and/or modify it
under the terms of the GNU GPL as published by the Free Software
Foundation, either version 3 of the License, or any later
version. {\texttt{ewN2HDECAY}} is distributed without any warranty. A
copy of the GNU GPL is included in the {\texttt{LICENSE.md}} file in
the root directory of {\texttt{ewN2HDECAY}}. 
\subsection{Download}
\label{sec:Download}
The latest version of the program package {\texttt{ewN2HDECAY}} can always be obtained from the url \url{https://github.com/marcel-krause/ewN2HDECAY}. The
user can either clone the repository or download the whole program as
a zip archive. The directory for the installation that is chosen by
the user will in the following be referred to as
{\texttt{\$ewN2HDECAY}}. The main folder of {\texttt{ewN2HDECAY}} 
contains several subfolders: 

\begin{description}
\item[{\texttt{BuildingBlocks}}] Here the analytic EW
   one-loop corrections for all considered decays can be found, as well as the
   CTs and real corrections needed for the UV and IR finiteness of the decay widths. 
\item[{\texttt{Documentation}}] Contains this documentation. 
\item[{\texttt{N2HDECAY}}] This subfolder contains a modified version
   of
   {\texttt{N2HDECAY}} \cite{DJOUADI199856,Djouadi:2018xqq,
     Muhlleitner:2016mzt}. It computes the LO and (where applicable) QCD
   corrected decay widths as well as off-shell
   decay widths and the loop-induced decay widths into gluon and
   photon pair final states and into $Z\gamma$. {\tt N2HDECAY} also
   computes the branching ratios.
 \item[{\texttt{Input}}] Here, at least one
   or more input files which shall be used for the computation are
   stored. In Sec.\,\ref{sec:InputFileFormat} the format of the input
   file is explained. In the Github repository, we provide the exemplary input file depicted in App.\,\ref{sec:AppendixInputFile} in the  folder {\texttt{Input}}. 
\item[{\texttt{Results}}] In this subfolder the results of a successful run of
  {\texttt{ewN2HDECAY}} are stored as output files. They have the same name as the
  corresponding input files in the {\texttt{Input}} folder, but with
  the file extension {\texttt{.in}} replaced by {\texttt{.out}} and a
  suffix ``\_BR'' and ``\_EW'' for the branching ratios and
  EW partial decay widths, respectively. In the Github
  repository, we provide the exemplary output files which are printed in App.\,\ref{sec:AppendixOutputFile} in the folder {\texttt{Results}}. 
\end{description}

The main folder {\texttt{\$ewN2HDECAY}} also contains several
files: 
\begin{description}
\item[{\texttt{ewN2HDECAY.py}}] This is the main program file of
  {\texttt{ewN2HDECAY}}. It serves as a wrapper file that calls 
  {\texttt{N2HDECAY}} for the conversion of the charm and bottom quark
  masses from the $\overline{\text{MS}}$ input values to the
  corresponding OS values and for the computation of the LO widths, QCD
  corrections, off-shell and loop-induced decays and the branching
  ratios. It also calls {\texttt{electroweakCorrections}} for the calculation of
  the EW one-loop corrections.
\item[{\texttt{Changelog.md}}] Documents all changes made in
  the {\texttt{ewN2HDECAY}} since version {\texttt{1.0.0}}. 
\item[{\texttt{CommonFunctions.py}}] A library of
  functions frequently used in the
  different files of the program {\texttt{ewN2HDECAY}}.
\item[{\texttt{Config.py}}] Main configuration file. In case 
  {\texttt{LoopTools}} is not installed automatically by the installer
  of {\texttt{ewN2HDECAY}}, the paths to the {\texttt{LoopTools}}
  executables and libraries have to be set manually in this file.
\item[{\texttt{constants.F90}}] Library for all constants
  used in {\texttt{ewN2HDECAY}}. 
\item[{\texttt{counterterms.F90}}] Here all fundamental CTs
  necessary for the EW one-loop renormalization of the Higgs boson 
  decays are defined. These CTs require the analytic results
  saved in the {\texttt{BuildingBlocks}} subfolder. 
\item[{\texttt{electroweakCorrections.F90}}] Main file for the
    calculation of the EW one-loop corrections to the Higgs boson 
    decays. The EW one-loop corrections to the decay
    widths are combined with the necessary CTs and real corrections
    for the EW contributions to the tree-level decay widths that 
    are then combined with the QCD corrections in
    {\texttt{N2HDECAY}}. 
\item[{\texttt{getParameters.F90}}] Routine to 
  read in the input values
    given by the user in the input files that are needed by
    {\texttt{ewN2HDECAY}}.  
\item[{\texttt{LICENSE.md}}] Contains the full GNU General Public
  License ({\texttt{GNU GPL-3.0-or-later}}) agreement. 
\item[{\texttt{README.md}}] Provides an overview over basic
  information about the program as well as a quick-start guide. 
\item[{\texttt{setup.py}}] Main setup and installation file of
    {\texttt{ewN2HDECAY}}. For a guided installation, this file should be
    called after downloading the program. 
\end{description}
\subsection{Installation}
\label{sec:Installation}
For the installation of {\texttt{ewN2HDECAY}}, we recommend to use the
automatic installation script {\texttt{setup.py}} which is part of the
{\texttt{ewN2HDECAY}} repository and which guides the user through the
installation. If {\texttt{ewN2HDECAY}} is installed under
{\texttt{Windows}}, the user should first install {\texttt{Cygwin}}
and check whether the path to the {\texttt{Cygwin}} executable in line
36 of the configuration file {\texttt{\$ewN2HDECAY/Config.py}} is
correct and modify it, if necessary. In order to use the automatic
installation script, the user opens a terminal, navigates to the main
folder {\texttt{\$ewN2HDECAY}} and executes the following command: 
\begin{lstlisting}[numbers=none,language=bash,frame=single,backgroundcolor=\color{mygray}]
python setup.py
\end{lstlisting}
First, the script asks whether the {\texttt{LoopTools}} version
specified in line 37 of {\texttt{\$2HDECAY/Config.py}} should
be downloaded and installed. By typing {\texttt{y}}, the specified
version of {\texttt{LoopTools}} is automatically installed in a
subdirectory of {\texttt{ewN2HDECAY}}. For a more detailed description
of the installation of the program, we refer to
\cite{HAHN1999153}. The installation of {\texttt{LoopTools}} is
optional if the user already has a current working version of the
program on the system. The installation of
{\texttt{LoopTools}} can then be skipped. However, in this case the lines
33-35 of the configuration file {\texttt{\$2HDECAY/Config.py}} need to
be modified in order to specify the path to the {\texttt{LoopTools}}
executables and libraries. Moreover, line 32 has to be set to  
\begin{lstlisting}[language=Python,frame=single,backgroundcolor=\color{mygray},numbers=none]
useRelativeLoopToolsPath = False
\end{lstlisting}
This step is crucial if {\texttt{LoopTools}} is not installed
automatically with the install script, since otherwise the paths to
the {\texttt{LoopTools}} libraries are incorrectly set in the makefile
of {\texttt{ewN2HDECAY}} and the make process of the program will
fail. 

In the next step of the installation, the script asks the user whether
the makefile and the file {\texttt{electroweakCorrections.F90}} should
be created and whether the program should be compiled, which should be
responded to by typing {\texttt{y}}. The script then executes the
makefile and the program is compiled. The make process may take
several minutes to finish. Finally, the script provides a convenient
way to 'make clean' the installation, which is optional.  

After the installation, the user can type
\begin{lstlisting}[numbers=none,language=bash,frame=single,backgroundcolor=\color{mygray}]
python ewN2HDECAY.py
\end{lstlisting}
into the terminal for a quick check whether the installation was
successful. The exemplary input file provided in the
{\texttt{2HDECAY}} repository is then used for the calculation of the
partial decay widths and branching ratios and the exemplary output
files provided in the {\texttt{\$2HDECAY/Results}} subdirectory should
be reproduced. If the program terminates without printing an error to
the terminal, the installation was successful. 

\subsection{Input File Format}
\label{sec:InputFileFormat}
\begin{table}[tb]
\centering
  \begin{tabular}{ c c c c }
    \hline
    Line & Input name & Allowed values and meaning \\ \hline
    \makecell[tc]{6} & \makecell[tc]{{\texttt{OMIT ELW2}}} &
\makecell[tl]{0: electroweak corrections (N2HDM) are calculated \\ 1: electroweak
    corrections (N2HDM) are neglected} \\  
    \makecell[tc]{10} & \makecell[tc]{{\texttt{N2HDM}}} &
\makecell[tl]{0: considered model is not the N2HDM \\ 1: considered
    model is the N2HDM } \\ 
    \makecell[tc]{58} & \makecell[tc]{{\texttt{TYPE}}} &
\makecell[tl]{1: N2HDM type I \\ 2: N2HDM type II \\ 
3: N2HDM lepton-specific \\ 4: N2HDM flipped} \\ 
    \makecell[tc]{76} & \makecell[tc]{{\texttt{RENSCHEM}}} &
\makecell[tl]{0: all renormalization schemes are calculated \\ 1-10: only the chosen scheme ({\it cf.}~Tab.\,\ref{tab:2HDECAYImplementedSchemes}) is calculated} \\ 
    \makecell[tc]{77} & \makecell[tc]{ {\texttt{REFSCHEM}} } &
\makecell[tl]{ 1-10: the input values of
    $\alpha _i$, $\beta$, $m_{12}^2$ and $v_S$
    (\textit{cf.}\,Tab.\,\ref{tab:2HDECAYInputValues}) are given in
    the \\ \hspace*{0.35cm} chosen reference
    scheme and at the scale $\mu _R$ given by {\texttt{INSCALE}} in 
\\ \hspace*{0.35cm} case of $\overline{\mbox{MS}}$
    parameters; the values of $\alpha _i$, $\beta$, $m_{12}^2$ and $v_S$ in all other 
\\ \hspace*{0.35cm} schemes and 
    at the scale $\mu_{\text{out}}$ at which the decays are calculated,
\\ \hspace*{0.35cm} are evaluated using
    Eqs.~(\ref{eq:scalechange}) and (\ref{eq:convertedParameterValues}) } \\ 
    \hline
  \end{tabular}
    \caption{Set of input parameters for the basic control of
      {\texttt{ewN2HDECAY}}. The first column depicts the line number
      at which the input value is specified in the input file. For the
      calculation of the one-loop EW corrections in the N2HDM, the
      parameter {\texttt{OMIT ELW2}} has to be set to 0. In this case,
      the value of {\texttt{N2HDM}} is set to 1 automatically
      internally and the provided input value is ignored. All input
      values presented in this table have to be entered as integer
      values.}   
   \label{tab:2HDECAYControlInputs} 
\end{table}
The format of the input file of {\texttt{ewN2HDECAY}} is directly
adopted from the input file format of {\texttt{N2HDECAY}}
\cite{Muhlleitner:2016mzt}, which again is adopted from the format of
{\texttt{HDECAY}} \cite{DJOUADI199856, Djouadi:2018xqq}. In comparison
to the {\texttt{N2HDECAY}} input file, minor modifications where
implemented to account for the EW corrections. In the input file, two
classes of input parameters are provided. The first class are the
integer-valued input parameters needed for the basic control of
{\texttt{ewN2HDECAY}}, as given in
Tab.\,\ref{tab:2HDECAYControlInputs}. The second class of input
parameters are the values of the independent physical input
parameters, provided in {\texttt{FORTRAN}} double-precision format, as
shown in Tab.\,\ref{tab:2HDECAYInputValues}. Since the input file
format of {\texttt{ewN2HDECAY}} is mostly equivalent to the one of
{\texttt{2HDECAY}} (apart from the extended scalar sector), we refer
to \cite{Krause:2018wmo} for an in-depth explanation of the meaning of
the input parameters. However, we want to emphasize here again that
the input parameters {\texttt{M\_12\^{}2}} and {\texttt{V\_SING}},
corresponding to the values of the $\overline{\text{MS}}$ parameters
$m_{12}^2$ and $v_S$, respectively, as well as {\texttt{alpha1}},
{\texttt{alpha2}}, {\texttt{alpha3}} and {\texttt{TGBET2HDM}},
corresponding to the values of $\alpha _i$ ($i=1,2,3$) and $\tan
\beta$ (in case that they are $\overline{\text{MS}}$ renormalized) all
explicitly depend on the input renormalization scale $\mu _R$, given
by {\texttt{INSCALE}}. This scale can be set to an arbitrary
double-precision value by the user. The EW one-loop corrected decay
widths are evaluated at the output scale $\mu _\text{out}$, provided
by {\texttt{OUTSCALE}}, which can be set to either a global fixed
value or to the mass of the decaying particle by setting
{\texttt{OUTSCALE=MIN}}. In case that both scales differ, all
$\overline{\text{MS}}$ parameters are converted from $\mu _R$ to $\mu
_\text{out}$ by means of \eqref{eq:scalechange}. Moreover, if the
reference renormalization scheme given by {\texttt{REFSCHEM}} is
different from the renormalization scheme {\texttt{RENSCHEM}} within
which the partial decay widths are calculated, then the values of the
scalar mixing angles are automatically converted from one scheme to
the other by means of \eqref{eq:convertedParameterValues}. 
\begin{table}[h]
\centering
  \begin{tabular}{ c c c c }
    \hline
    Line & Input name & Name in Sec.\,\ref{sec:EWQCDN2HDMMain} & Allowed values and meaning \\ \hline
    \makecell[tc]{19} & \makecell[tc]{{\texttt{ALS(MZ)}}} & $\alpha_s (m_Z)$ & \makecell[tl]{strong coupling constant (at $m_Z$)} \\ 
    \makecell[tc]{20} & \makecell[tc]{{\texttt{MSBAR(2)}}} & $m_s (2\,\text{GeV})$ & \makecell[tl]{$s$-quark $\overline{\text{MS}}$ mass at 2 GeV in GeV} \\ 
    \makecell[tc]{21} & \makecell[tc]{{\texttt{MCBAR(3)}}} & $m_c (3\,\text{GeV})$ & \makecell[tl]{$c$-quark $\overline{\text{MS}}$ mass at 3 GeV in GeV} \\ 
    \makecell[tc]{22} & \makecell[tc]{{\texttt{MBBAR(MB)}}} & $m_b (m_b)$ & \makecell[tl]{$b$-quark $\overline{\text{MS}}$ mass at $m_b$ in GeV} \\ 
    \makecell[tc]{23} & \makecell[tc]{{\texttt{MT}}} & $m_t$ & \makecell[tl]{$t$-quark pole mass in GeV} \\ 
    \makecell[tc]{24} & \makecell[tc]{{\texttt{MTAU}}} & $m_\tau $ & \makecell[tl]{$\tau$-lepton pole mass in GeV} \\ 
    \makecell[tc]{25} & \makecell[tc]{{\texttt{MMUON}}} & $m_\mu $ & \makecell[tl]{$\mu $-lepton pole mass in GeV} \\ 
    \makecell[tc]{26} & \makecell[tc]{{\texttt{1/ALPHA}}} & $\alpha _\text{em} ^{-1} (0)$ & \makecell[tl]{inverse fine-structure constant (Thomson limit)} \\ 
    \makecell[tc]{27} & \makecell[tc]{{\texttt{ALPHAMZ}}} & $\alpha _\text{em} (m_Z)$ & \makecell[tl]{fine-structure constant (at $m_Z$)} \\ 
    \makecell[tc]{30} & \makecell[tc]{{\texttt{GAMW}}} & $\Gamma _W$ & \makecell[tl]{partial decay width of the $W$ boson } \\ 
    \makecell[tc]{31} & \makecell[tc]{{\texttt{GAMZ}}} & $\Gamma _Z$ & \makecell[tl]{partial decay width of the $Z$ boson } \\ 
    \makecell[tc]{32} & \makecell[tc]{{\texttt{MZ}}} & $m_Z$ &
\makecell[tl]{$Z$ boson on-shell mass in GeV} \\ 
    \makecell[tc]{33} & \makecell[tc]{{\texttt{MW}}} & $m_W$ &
\makecell[tl]{$W$ boson on-shell mass in GeV} \\ 
    \makecell[tc]{34-42} & \makecell[tc]{{\texttt{Vij}}} & $V_{ij}$ & \makecell[tl]{CKM matrix elements ($i\in \{ u,c,t\}$ , $j\in \{ d,s,b\} $) } \\ 
    \makecell[tc]{60} & \makecell[tc]{{\texttt{TGBET2HDM}}} & $t_\beta $ & \makecell[tl]{ratio of the VEVs in the 2HDM} \\ 
    \makecell[tc]{61} & \makecell[tc]{{\texttt{M\_12\textasciicircum
2}}} & $m_{12}^2 $ & \makecell[tl]{squared soft-$\mathbb{Z}_2$-breaking scale in GeV$^2$} \\ 
    \makecell[tc]{66} & \makecell[tc]{{\texttt{MHA}}} & $m_A $ &
\makecell[tl]{CP-odd Higgs boson mass in GeV} \\
    \makecell[tc]{67} & \makecell[tc]{{\texttt{MH+-}}} & $m_{H^\pm } $
& \makecell[tl]{charged Higgs boson mass in GeV} \\
    \makecell[tc]{78} & \makecell[tc]{{\texttt{INSCALE}}} & $\mu_R $
 & \makecell[tl]{renormalization scale for $\overline{\text{MS}}$ inputs in GeV} \\ 
    \makecell[tc]{79} & \makecell[tc]{ {\texttt{OUTSCALE}} } & $\mu_\text{out} $
 & \makecell[tl]{ renormalization scale for the evaluation of the } \\ 
      &   &   & \makecell[tl]{ partial decay widths in GeV or in terms of {\texttt{MIN}} } \\ 
    \makecell[tc]{80} & \makecell[tc]{{\texttt{MH1}}} & $m_{H_1} $ &
\makecell[tl]{mass of the CP-even Higgs boson $H_1$ in GeV} \\
    \makecell[tc]{81} & \makecell[tc]{{\texttt{MH2}}} & $m_{H_2} $ &
\makecell[tl]{mass of the CP-even Higgs boson $H_2$ in GeV} \\
    \makecell[tc]{82} & \makecell[tc]{{\texttt{MH3}}} & $m_{H_3} $ &
\makecell[tl]{mass of the CP-even Higgs boson $H_3$ in GeV} \\
    \makecell[tc]{83} & \makecell[tc]{{\texttt{alpha1}}} & $\alpha _1 $ & \makecell[tl]{CP-even Higgs mixing angle $\alpha _1$ in radians} \\ 
    \makecell[tc]{84} & \makecell[tc]{{\texttt{alpha2}}} & $\alpha _2 $ & \makecell[tl]{CP-even Higgs mixing angle $\alpha _2$ in radians} \\ 
    \makecell[tc]{85} & \makecell[tc]{{\texttt{alpha3}}} & $\alpha _3 $ & \makecell[tl]{CP-even Higgs mixing angle $\alpha _3$ in radians} \\ 
    \hline
  \end{tabular}
    \caption{All input parameters necessary for running {\texttt{ewN2HDECAY}}. The number in
      the first column denotes the line in the input file at which the
      input value is specified.  
      Almost all input parameters are entered as integers. Exceptions
      are the renormalization scales {\texttt{INSCALE}}, entered as a
      double-precision number, and {\texttt{OUTSCALE}}, entered as
      either a double-precision number or in terms of the mass
      {\texttt{MIN}} of the decaying particle. Note
        that the 2HDM values $t_\beta$, $m_{12}^2$, $m_A$ and
        $m_{H^\pm}$ are used as N2HDM input parameters if {\tt
          N2HDM} is set to 1 in the input file.}  
   \label{tab:2HDECAYInputValues}
\end{table}
The user has the possibility to store as many input files in the input
file subdirectory {\texttt{\$2HDECAY/Input}} as wanted. The input
files are allowed to have arbitrary non-empty filenames and 
filename extensions. After a successful run, all output files are
stored in the output file subdirectory {\texttt{\$2HDECAY/Results}}
under the same name as the corresponding input file, but with the
filename extension replaced by {\texttt{.out}}. For each input file
that is used for the calculation, two corresponding output files are
created. One output file, indicated by the filename suffix '\_BR',
contains the calculated branching ratios, while the other, indicated
by the filename suffix '\_EW', contains the electroweak partial decay
widths, {\it cf.}~Subsection~\ref{sec:n2hdecaylink}.
\begin{table}[tb]
\centering
  \begin{tabular}{ c c c c c }
    \hline
    Input ID & Tadpole scheme & $\delta \alpha _i$ & $\delta \beta$ & Gauge-par.-indep. $\Gamma$ \\ \hline
    \makecell[tc]{1} & \makecell[tc]{standard} & \makecell[tc]{KOSY} & \makecell[tc]{KOSY (odd)} & \makecell[tc]{\xmark } \\ 
    \makecell[tc]{2} & \makecell[tc]{standard} & \makecell[tc]{KOSY} & \makecell[tc]{KOSY (charged)} & \makecell[tc]{\xmark } \\ 
    \makecell[tc]{3} & \makecell[tc]{alternative (FJ)} & \makecell[tc]{KOSY} & \makecell[tc]{KOSY (odd)} & \makecell[tc]{\xmark } \\ 
    \makecell[tc]{4} & \makecell[tc]{alternative (FJ)} & \makecell[tc]{KOSY} & \makecell[tc]{KOSY (charged)} & \makecell[tc]{\xmark } \\ 
    \makecell[tc]{5} & \makecell[tc]{alternative (FJ)} & \makecell[tc]{$p_{*}$-pinched} & \makecell[tc]{$p_{*}$-pinched (odd)} & \makecell[tc]{\cmark } \\ 
    \makecell[tc]{6} & \makecell[tc]{alternative (FJ)} & \makecell[tc]{$p_{*}$-pinched} & \makecell[tc]{$p_{*}$-pinched (charged)} & \makecell[tc]{\cmark } \\ 
    \makecell[tc]{7} & \makecell[tc]{alternative (FJ)} & \makecell[tc]{OS-pinched} & \makecell[tc]{OS-pinched (odd)} & \makecell[tc]{\cmark } \\ 
    \makecell[tc]{8} & \makecell[tc]{alternative (FJ)} & \makecell[tc]{OS-pinched} & \makecell[tc]{OS-pinched (charged)} & \makecell[tc]{\cmark } \\ 
    \makecell[tc]{9} & \makecell[tc]{standard} & \makecell[tc]{ $\overline{\text{MS}}$ } & \makecell[tc]{ $\overline{\text{MS}}$ } & \makecell[tc]{\xmark } \\ 
    \makecell[tc]{10} & \makecell[tc]{alternative (FJ)} & \makecell[tc]{ $\overline{\text{MS}}$ } & \makecell[tc]{ $\overline{\text{MS}}$ } & \makecell[tc]{\cmark } \\ 
    \hline
  \end{tabular}
    \caption{Overview over all renormalization schemes for the mixing
      angles $\alpha_i$ ($i=1,2,3$) and $\beta$ that are implemented in
      {\texttt{ewN2HDECAY}}. By setting {\texttt{RENSCHEM}} in the input
      file, {\it cf.}~Tab.\,\ref{tab:2HDECAYControlInputs}, equal to the
      Input ID the corresponding renormalization scheme is chosen. In
      case of 0, however, the
      results for all renormalization schemes are given out. The
      definition of the CTs $\delta \alpha_i$ and $\delta \beta$ in each
      scheme is explained in
      Sec.\,\ref{sec:renormalizationMixingAngles}. The crosses and
      check marks in the column for gauge independence indicate
      whether the chosen scheme in general yields explicitly gauge-independent
      partial decay widths or not.} 
   \label{tab:2HDECAYImplementedSchemes}
\end{table}

\subsection{Structure of the Program}
\begin{figure}[h]
\centering
\includegraphics[width=13.8cm]{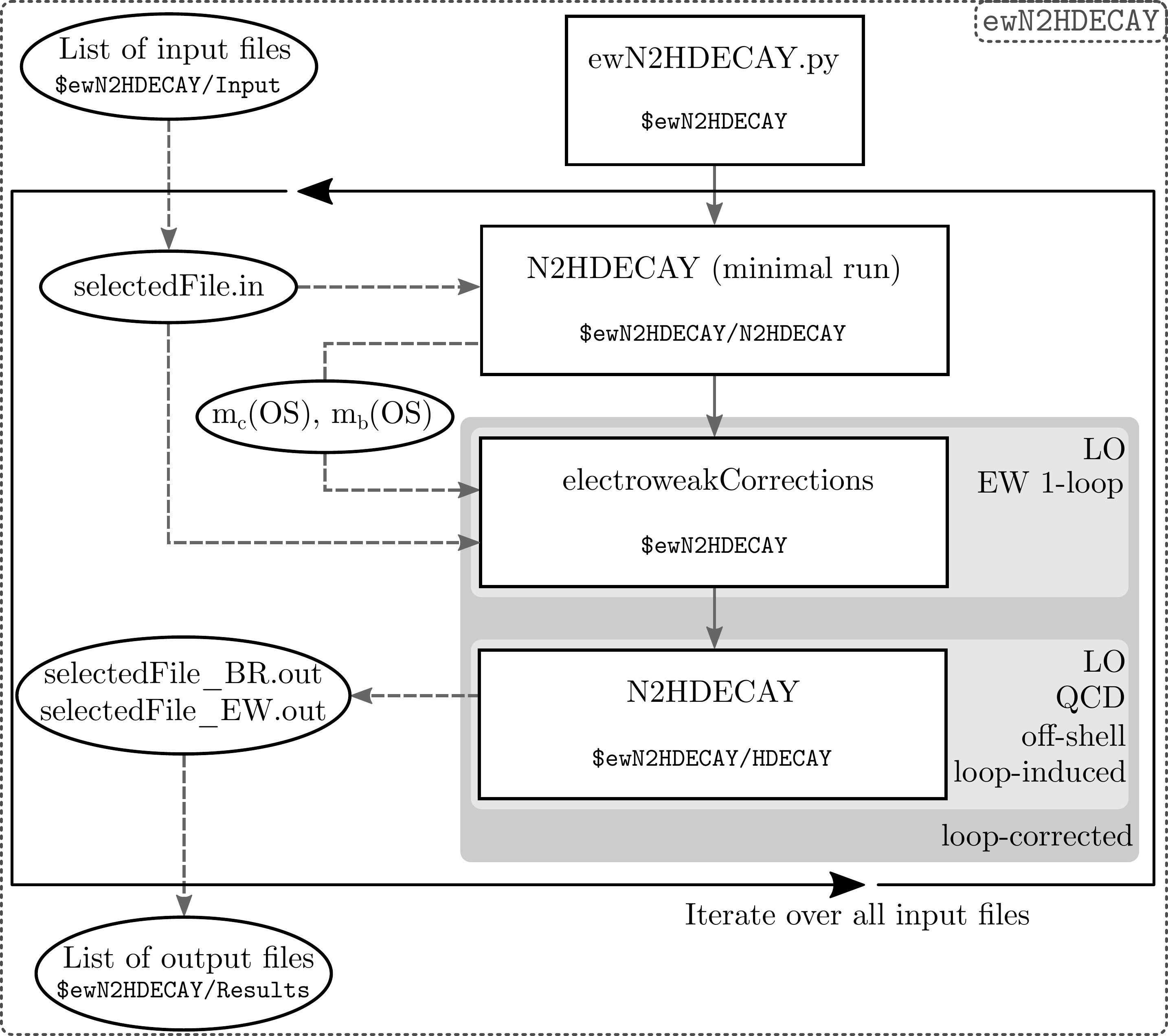}
\caption{Flowchart of {\texttt{ewN2HDECAY}}. The main wrapper file
  {\texttt{ewN2HDECAY.py}} generates a list of input files, provided by the
  user in the subfolder {\texttt{\$ewN2HDECAY/Input}}, and iterates over
  the list. For each selected input file in the list, the wrapper
  calls {\tt N2HDECAY} and the subprogram {\tt
    electroweakCorrections}. The computed branching ratios including
  the EW and QCD corrections as described in the text are written 
to the output file with suffix '\_BR', the calculated LO and NLO EW-corrected
partial decay widths are given out in the output file with suffix
'\_EW'.}
\label{fig:flowchartewN2HDECAY}
\end{figure}

The program {\texttt{ewN2HDECAY}} combines the EW one-loop corrections
to the partial decay widths with the state-of-the-art QCD corrections
already available in the tool {\texttt{N2HDECAY}}. In
\figref{fig:flowchartewN2HDECAY}, the work flow of the main wrapper
file of {\texttt{ewN2HDECAY}},
\textit{i.e.}\,{\texttt{ewN2HDECAY.py}}, is depicted.  
The wrapper file iterates over all given input files and first calls
{\texttt{N2HDECAY}} in a so-called minimal run in which the charm and
bottom masses are converted from their $\overline{\mbox{MS}}$ values
to the corresponding pole masses. These are then used together with
all other given input values to calculate the EW corrections to the
partial decay widths. In a second run of {\texttt{N2HDECAY}}, the LO
widths and QCD corrections to the decays are computed. Additionally,
the off-shell decay widths and the loop-induced decays to final-state
pairs of gluons or photons and $Z \gamma$ are calculated and the
branching ratios are evaluated. The results of these computations are
then consistently combined with the EW corrections as described in
Subsection~\ref{sec:n2hdecaylink}. This procedure is repeated for
every individual input file in the input file list. The work flow of
{\texttt{ewN2HDECAY}} is analogous to that of {\texttt{2HDECAY}}. For
a more detailed description, we refer to \cite{Krause:2018wmo}. 

\subsection{Usage}
Before running {\texttt{ewN2HDECAY}}, all input files need to be
stored in the input file subdirectory {\texttt{\$ewN2HDECAY/Input}}. The input files have to be
formatted exactly as described in Sec.\,\ref{sec:InputFileFormat},
otherwise, the program might either crash with a segmentation error or
input values are read in incorrectly. For convenience, the user may
use the exemplary input file printed in
App.\,\ref{sec:AppendixInputFile} which is part of the
{\texttt{ewN2HDECAY}} repository as a template to generate own input files.

In each run of {\texttt{ewN2HDECAY}}, the output file subfolder
{\texttt{\$ewN2HDECAY/Results}} is emptied to make space for the
output files generated in a new run. The user is therefore advised
to check the output file folder for any output files of a previous run
before starting a new run and to create backups of them, if
necessary. 

In order to start a run of {\texttt{ewN2HDECAY}}, the user opens a
terminal, navigates to the main {\texttt{\$ewN2HDECAY}} folder and
executes the following command:
\begin{lstlisting}[numbers=none,language=bash,frame=single,backgroundcolor=\color{mygray}]
python ewN2HDECAY.py
\end{lstlisting}
Provided that {\texttt{ewN2HDECAY}} was installed successfully and that
all input files stored in the subdirectory
{\texttt{\$ewN2HDECAY/Input}} have the correct format,
{\texttt{ewN2HDECAY}} now iterates over all input files and computes
the EW and/or QCD corrections as indicated by the flowchart in
\figref{fig:flowchartewN2HDECAY}. Intermediate results and some 
additional information are printed on the terminal. After a successful
run, {\texttt{ewN2HDECAY}} terminates with no errors printed on the
terminal and the output files are stored in the subdirectory
{\texttt{\$ewN2HDECAY/Results}}. 

\subsection{Output File Format}
\label{sec:OutputFileFormat}
After a successful run, {\texttt{ewN2HDECAY}} creates two output files
for each individual input file, indicated in their filenames by the
suffixes '\_QCD' and '\_EW' for the branching ratios and electroweak
partial decay widths, respectively. Exemplary output files are printed
in App.\,\ref{sec:AppendixOutputFile}. The output file format is 
a modified SLHA format\footnote{The original SLHA output format
  \cite{Skands:2003cj,Allanach:2008qq,Mahmoudi:2010iz} has only been
  designed for supersymmetric models. For {\texttt{ewN2HDECAY}}, we
  modified the SLHA format to account for the EW corrections
  calculated in the N2HDM.}. The format of the output files is
completely analogous to the format of the output files in
{\texttt{2HDECAY}}. For a detailed description, we refer to
\cite{Krause:2018wmo}. 

\section{Summary}
\label{sec:summary}
We have presented the program package {\texttt{ewN2HDECAY}} for the 
calculation of the EW corrections to the Higgs boson decays in the
N2HDM. The program allows for the computation of NLO EW corrections to
the two-body on-shell decays of all N2HDM Higgs bosons that are not
loop-induced. For the scalar mixing angles $\alpha _i$ ($i=1,2,3$) and
$\beta$ of the N2HDM, 10 different renormalization schemes are
implemented. The other parameters of the N2HDM necessary for the
computation of the EW corrections are renormalized in an OS scheme,
except for the soft-$\mathbb{Z}_2$-breaking scale $m_{12}^2$ and the
singlet VEV $v_S$ which are $\overline{\text{MS}}$ renormalized. The
EW corrections are consistently combined with the
state-of-the-art QCD corrections that are obtained from
{\texttt{N2HDECAY}} and the EW\&QCD-corrected 
total decay widths and branching ratios are given out in an
SLHA-inspired output file format. Additionally, {\texttt{ewN2HDECAY}}
provides a separate SLHA-inspired output of the LO and NLO EW partial
decay widths to all OS non-loop-induced N2HDM Higgs decays. 
The implementation of 10 different renormalization schemes for the
scalar mixing angles and a routine to automatically convert the scalar
mixing angles from one scheme to another, allows to
compare NLO partial decay widths calculated within these different
schemes. The $\overline{\text{MS}}$ 
input parameters are moreover evolved from a user-given input
renormalization scale $\mu _R$ to a user-given output scale $\mu
_\text{out}$ at which the partial decay widths are evaluated. 
Thus the remaining theoretical uncertainty due to missing higher-order
corrections can be estimated from a scale change or the change of the
renormalization scheme. Because of its fast numerical computation, our
new program {\texttt{ewN2HDECAY}} allows for efficient
phenomenological studies of the N2HDM Higgs 
sector at high precision required for the analysis of indirect new physics
effects in the Higgs sector and for the identification of the
underlying model in case additional Higgs bosons are discovered.  

\subsection*{Acknowledgments}
We thank Jonas Wittbrodt, one of the authors of {\tt N2HDECAY}, for
agreeing to use the code as basis for the QCD corrected decay widths. We thank David Lopez-Val for independently cross-checking some of the
analytic results derived for this work. MK and MM acknowledge
financial support from the DFG project “Precision 
Calculations in the Higgs Sector - Paving the Way to the New Physics
Landscape” (ID: MU 3138/1-1). 

\begin{appendix}

\section{Exemplary Input File}
\label{sec:AppendixInputFile}
We present an exemplary input file
{\texttt{ewn2hdecay.in}} which is also included in the program subfolder
{\texttt{\$ewN2HDECAY/Input}} in the {\texttt{ewN2HDECAY}} repository. The
first integer in each line represents the line number and is not part
of the actual input file, but given here for convenience. The
role of the various input parameters has been specified in
Sec.\,\ref{sec:InputFileFormat}. With respect to the input file
format of the unmodified {\texttt{N2HDECAY}}
program\cite{Muhlleitner:2016mzt}, the lines 6, 27, 29, and 76-79 are
new, but the remainder of the input file format is unchanged. Note
that the value {\texttt{GFCALC}} in the input 
file is overwritten by the program and thus not an input value that
is provided by the user, but it is calculated by {\texttt{ewN2HDECAY}}
internally. The sample N2HDM parameter point has been checked
  to be in accordance with all relevant theoretical and experimental
  constraints. Thus, it features an SM-like Higgs boson with a mass of 
  125.09~GeV \cite{Aad:2015zhl} which is given by the lightest CP-even neutral Higgs
  boson $H_1$. For details on the applied constraints, we refer to
  Ref.~\cite{Muhlleitner:2016mzt}. Note that we understood all
  running $\overline{\mbox{MS}}$ input parameters to be given by the
  mass of the SM-like Higgs boson by specifying the value {\tt
    INSCALE} accordingly. The EW-corrected decays on the other hand
  are calculated at the scale of the decaying particle, {\it
    i.e.}~{\tt OUTSCALE=MIN}.

\begin{lstlisting}
SLHAIN   = 1
SLHAOUT  = 1
COUPVAR  = 1
HIGGS    = 5
OMIT ELW = 1
OMIT ELW2= 0
SM4      = 0
FERMPHOB = 0
2HDM     = 1
N2HDM    = 1
MODEL    = 1
TGBET    = 5.07403e+01
MABEG    = 4.67967e+02
MAEND    = 4.67967e+02
NMA      = 1
********************* hMSSM (MODEL = 10) *********************************
MHL      = 125.D0
**************************************************************************
ALS(MZ)  = 1.18000e-01
MSBAR(2) = 9.50000e-02
MCBAR(3) = 0.98600e+00
MBBAR(MB)= 4.18000e+00
MT       = 1.73200e+02
MTAU     = 1.77682e+00
MMUON    = 1.056583715e-01
1/ALPHA  = 1.37036e+02
ALPHAMZ  = 7.754222173973729e-03
GF       = 1.1663787e-05
GFCALC   = 0.000000000
GAMW     = 2.08500e+00
GAMZ     = 2.49520e+00
MZ       = 9.11876e+01
MW       = 8.0385e+01
VTB      = 9.9910e-01
VTS      = 4.040e-02
VTD      = 8.67e-03
VCB      = 4.12e-02
VCS      = 9.7344e-01
VCD      = 2.252e-01
VUB      = 3.51e-03
VUS      = 2.2534e-01
VUD      = 9.7427e-01
********************* 4TH GENERATION *************************************
  SCENARIO FOR ELW. CORRECTIONS TO H -> GG (EVERYTHING IN GEV):
  GG_ELW = 1: MTP = 500    MBP = 450    MNUP = 375    MEP = 450
  GG_ELW = 2: MBP = MNUP = MEP = 600    MTP = MBP+50*(1+LOG(M_H/115)/5)

GG_ELW   = 1
MTP      = 500.D0
MBP      = 450.D0
MNUP     = 375.D0
MEP      = 450.D0
************************** 2 Higgs Doublet Model *************************
  TYPE: 1 (I), 2 (II), 3 (lepton-specific), 4 (flipped)
  PARAM: 1 (masses), 2 (lambda_i)

PARAM    = 1
TYPE     = 2
********************
TGBET2HDM= 1.32331D0
M_12^2   = 182778.0D0
******************** PARAM=1:
ALPHA_H  = 10.D0
MHL      = 10.D0
MHH      = 10.D0
MHA      = 722.000D0
MH+-     = 758.918D0
******************** PARAM=2:
LAMBDA1  = 0D0
LAMBDA2  = 0D0
LAMBDA3  = 0D0
LAMBDA4  = 0D0
LAMBDA5  = 0D0
**************************** N2HDM ***********************************
*** needs TYPE, TGBET2HDM, M12^2, MHA and MH+- from the 2HDM block ***
RENSCHEM = 7
REFSCHEM = 5
INSCALE  = 125.09D0
OUTSCALE = MIN
MH1      = 125.09D0
MH2      = 286.094D0
MH3      = 648.564D0
alpha1   = 0.909079D0
alpha2   = -0.155397D0
alpha3   = -1.54459D0
V_SING   = 2440.84D0
**************************************************************************
SUSYSCALE= 1000.D0
MU       = 1000.D0
M2       = 1000.D0
MGLUINO  = 1000.D0
MSL1     = 1000.D0
MER1     = 1000.D0
MQL1     = 1000.D0
MUR1     = 1000.D0
MDR1     = 1000.D0
MSL      = 1000.D0
MER      = 1000.D0
MSQ      = 1000.D0
MUR      = 1000.D0
MDR      = 1000.D0
AL       = 1000.D0
AU       = 1000.D0
AD       = 1000.D0
ON-SHELL = 0
ON-SH-WZ = 0
IPOLE    = 0
OFF-SUSY = 0
INDIDEC  = 0
NF-GG    = 5
IGOLD    = 0
MPLANCK  = 2.4D18
MGOLD    = 1.D-13
******************* VARIATION OF HIGGS COUPLINGS *************************
ELWK     = 0
CW       = 1.D0
CZ       = 1.D0
Ctau     = 1.D0
Cmu      = 1.D0
Ct       = 1.D0
Cb       = 1.D0
Cc       = 1.D0
Cs       = 1.D0
Cgaga    = 0.D0
Cgg      = 0.D0
CZga     = 0.D0
********************* 4TH GENERATION *************************************
Ctp      = 1.D0
Cbp      = 1.D0
Cnup     = 1.D0
Cep      = 1.D0
\end{lstlisting}

\section{Exemplary Output Files}
\label{sec:AppendixOutputFile}
We give here the exemplary output files
  {\texttt{ewn2hdecay\_BR.out}} and {\texttt{ewn2hdecay\_EW.out}} that
  have been generated from the sample input file
    {\texttt{ewn2hdecay.in}} and are included in the subfolder
    {\texttt{\$ewN2HDECAY/Results}} in the 
  {\texttt{ewN2HDECAY}} repository. The suffixes ``\_BR'' and ``\_EW''
  stand for the branching ratios and EW partial decay widths, given
  out in the respective files. Again, the first integer in each line represents the line
number and is not part of the actual output file, but printed here for
convenience. The output file format is explained in detail in
Sec.\,\ref{sec:OutputFileFormat}. The exemplary output file was
generated for a specific choice of the renormalization scheme, namely~we
have set {\texttt{RENSCHEM = 7}} in line 76 of the input file,
{\it cf.}~App.\,\ref{sec:AppendixInputFile}. For {\texttt{RENSCHEM = 0}},
the output file becomes considerably longer, since the EW
corrections are then calculated for all 10 implemented renormalization
schemes. As reference scheme we chose the renormalization scheme
number 5, {\it cf.}~Tab.~\ref{tab:2HDECAYImplementedSchemes} for its definition.

\subsection{Exemplary Output File for the Branching Ratios}
The exemplary output file {\texttt{ewn2hdecay\_BR.out}} contains the
branching ratios without and with the electroweak corrections. The
content of the file is presented in the following. 
\begin{lstlisting}
#
BLOCK DCINFO  # Decay Program information
     1   ewN2HDECAY  # decay calculator
     2   1.0.0       # version number
#
BLOCK SMINPUTS  # Standard Model inputs
         2     1.19596488E-05   # G_F [GeV^-2]
         3     1.18000000E-01   # alpha_S(M_Z)^MSbar
         4     9.11876000E+01   # M_Z on-shell mass
         5     4.18000000E+00   # mb(mb)^MSbar
         6     1.73200000E+02   # mt pole mass
         7     1.77682000E+00   # mtau pole mass
         8     4.84141297E+00   # mb pole mass
         9     1.43141297E+00   # mc pole mass
        10     1.05658372E-01   # muon mass
        11     8.03850000E+01   # M_W on-shell mass
        12     2.08500000E+00   # W boson total width
        13     2.49520000E+00   # Z boson total width
#
BLOCK N2HDMINPUTS  # N2HDM inputs of reference scheme
         1                  2   # N2HDM type
         2     1.32331000E+00   # tan(beta)
         3     1.82778000E+05   # M_12^2
         4     1.25090000E+02   # M_H1
         5     2.86094000E+02   # M_H2
         6     6.48564000E+02   # M_H3
         7     7.22000000E+02   # M_A
         8     7.58918000E+02   # M_CH
         9     9.09079000E-01   # alpha1
        10    -1.55397000E-01   # alpha2
        11    -1.54459000E+00   # alpha3
        12     2.44084000E+03   # v_S
        13                  7   # renormalization scheme EW corrs
        14                  5   # reference ren. scheme EW corrs
        15     125.09D0         # ren. scale of MSbar parameters
        16     MIN              # ren. scale at which decay is evaluated
#
BLOCK VCKMIN  # CKM mixing
         1     9.99100000E-01   # V_tb
         2     4.04000000E-02   # V_ts
         3     8.67000000E-03   # V_td
         4     4.12000000E-02   # V_cb
         5     9.73440000E-01   # V_cs
         6     2.25200000E-01   # V_cd
         7     3.51000000E-03   # V_ub
         8     2.25340000E-01   # V_us
         9     9.74270000E-01   # V_ud
#
#            PDG    Width QCD Only
DECAY QCD    25     4.14310207E-03   # H1 decays with QCD corrections only
                       7                # Renormalization Scheme Number
       0.90516083E+00   # Corresponding mixing angle   alpha1
      -0.15530063E+00   # Corresponding mixing angle   alpha2
      -0.15445673E+01   # Corresponding mixing angle   alpha3
       0.13029492E+01   # Corresponding tan(beta)
       0.18277800E+06   # Corresponding m_12^2
       0.24408400E+04   # Corresponding v_S
#          BR         NDA      ID1       ID2
     5.98141907E-01    2           5        -5   # BR(H1 -> b       bb    )
     6.43400735E-02    2         -15        15   # BR(H1 -> tau+    tau-  )
     2.27785781E-04    2         -13        13   # BR(H1 -> mu+     mu-   )
     2.25766183E-04    2           3        -3   # BR(H1 -> s       sb    )
     2.79834641E-02    2           4        -4   # BR(H1 -> c       cb    )
     7.45745241E-02    2          21        21   # BR(H1 -> g       g     )
     2.21308229E-03    2          22        22   # BR(H1 -> gam     gam   )
     1.54082518E-03    2          22        23   # BR(H1 -> Z       gam   )
     2.05127755E-01    2          24       -24   # BR(H1 -> W+      W-    )
     2.56248172E-02    2          23        23   # BR(H1 -> Z       Z     )
#
#               PDG    Width QCD and EW
DECAY QCD&EW    25     4.04594148E-03   # H1 decays with QCD and EW corrections
                       7                # Renormalization Scheme Number
       0.90516083E+00   # Corresponding mixing angle   alpha1
      -0.15530063E+00   # Corresponding mixing angle   alpha2
      -0.15445673E+01   # Corresponding mixing angle   alpha3
       0.13029492E+01   # Corresponding tan(beta)
       0.18277800E+06   # Corresponding m_12^2
       0.24408400E+04   # Corresponding v_S
#          BR         NDA      ID1       ID2
     5.92535099E-01    2           5        -5   # BR(H1 -> b       bb      )
     6.29924711E-02    2         -15        15   # BR(H1 -> tau+    tau -   )
     2.18132813E-04    2         -13        13   # BR(H1 -> mu+     mu-     )
     2.25522035E-04    2           3        -3   # BR(H1 -> s       sb      )
     2.75253978E-02    2           4        -4   # BR(H1 -> c       cb      )
     7.63653816E-02    2          21        21   # BR(H1 -> g       g       )
     2.26622799E-03    2          22        22   # BR(H1 -> gam     gam     )
     1.57782707E-03    2          22        23   # BR(H1 -> Z       gam     )
     2.10053761E-01    2          24       -24   # BR(H1 -> W+      W-      )
     2.62401801E-02    2          23        23   # BR(H1 -> Z       Z       )
#
#            PDG    Width QCD Only
DECAY QCD    35     2.51908281E-01   # H2 decays with QCD corrections only
                       7                # Renormalization Scheme Number
       0.90516083E+00   # Corresponding mixing angle   alpha1
      -0.15530063E+00   # Corresponding mixing angle   alpha2
      -0.15445673E+01   # Corresponding mixing angle   alpha3
       0.13029492E+01   # Corresponding tan(beta)
       0.16715360E+06   # Corresponding m_12^2
       0.24408400E+04   # Corresponding v_S
#          BR         NDA      ID1       ID2
     6.94510276E-04    2           5        -5   # BR(H2 -> b       bb    )
     8.77457425E-05    2         -15        15   # BR(H2 -> tau+    tau-  )
     3.10346729E-07    2         -13        13   # BR(H2 -> mu+     mu-   )
     2.60430448E-07    2           3        -3   # BR(H2 -> s       sb    )
     1.65447031E-05    2           4        -4   # BR(H2 -> c       cb    )
     5.64810018E-06    2           6        -6   # BR(H2 -> t       tb    )
     3.17272086E-04    2          21        21   # BR(H2 -> g       g     )
     1.04971784E-05    2          22        22   # BR(H2 -> gam     gam   )
     4.57638077E-05    2          22        23   # BR(H2 -> Z       gam   )
     4.67693293E-01    2          24       -24   # BR(H2 -> W+      W-    )
     2.05903660E-01    2          23        23   # BR(H2 -> Z       Z     )
     3.25224495E-01    2          25        25   # BR(H2 -> H1      H1    )
#
#               PDG    Width QCD and EW
DECAY QCD&EW    35     2.87330365E-01   # H2 decays with QCD and EW correctis
                       7                # Renormalization Scheme Number
       0.90516083E+00   # Corresponding mixing angle   alpha1
      -0.15530063E+00   # Corresponding mixing angle   alpha2
      -0.15445673E+01   # Corresponding mixing angle   alpha3
       0.13029492E+01   # Corresponding tan(beta)
       0.16715360E+06   # Corresponding m_12^2
       0.24408400E+04   # Corresponding v_S
#          BR         NDA      ID1       ID2
     5.72230258E-04    2           5        -5   # BR(H2 -> b       bb      )
     7.18420289E-05    2         -15        15   # BR(H2 -> tau+    tau -   )
     2.48410041E-07    2         -13        13   # BR(H2 -> mu+     mu-     )
     2.18867293E-07    2           3        -3   # BR(H2 -> s       sb      )
     1.38893457E-05    2           4        -4   # BR(H2 -> c       cb      )
     4.95180244E-06    2           6        -6   # BR(H2 -> t       tb      )
     2.78158787E-04    2          21        21   # BR(H2 -> g       g       )
     9.20308634E-06    2          22        22   # BR(H2 -> gam     gam     )
     4.01220460E-05    2          22        23   # BR(H2 -> Z       gam     )
     4.13024816E-01    2          24       -24   # BR(H2 -> W+      W-      )
     1.81037672E-01    2          23        23   # BR(H2 -> Z       Z       )
     4.04946647E-01    2          25        25   # BR(H2 -> H1      H1      )
#            PDG    Width QCD Only
DECAY QCD    38     1.46796633E+01   # H3 decays with QCD corrections only
                       7                # Renormalization Scheme Number
       0.90516083E+00   # Corresponding mixing angle   alpha1
      -0.15530063E+00   # Corresponding mixing angle   alpha2
      -0.15445673E+01   # Corresponding mixing angle   alpha3
       0.13029492E+01   # Corresponding tan(beta)
       0.15169628E+06   # Corresponding m_12^2
       0.24408400E+04   # Corresponding v_S
#          BR         NDA      ID1       ID2
     1.05566345E-03    2           5        -5   # BR(H3 -> b       bb    )
     1.55539709E-04    2         -15        15   # BR(H3 -> tau+    tau-  )
     5.50024113E-07    2         -13        13   # BR(H3 -> mu+     mu-   )
     3.86057017E-07    2           3        -3   # BR(H3 -> s       sb    )
     1.93693674E-05    2           4        -4   # BR(H3 -> c       cb    )
     9.92185598E-01    2           6        -6   # BR(H3 -> t       tb    )
     2.31958632E-03    2          21        21   # BR(H3 -> g       g     )
     6.71149392E-06    2          22        22   # BR(H3 -> gam     gam   )
     1.86917571E-06    2          22        23   # BR(H3 -> Z       gam   )
     1.29809388E-03    2          24       -24   # BR(H3 -> W+      W-    )
     6.32155300E-04    2          23        23   # BR(H3 -> Z       Z     )
     1.79993258E-03    2          25        25   # BR(H3 -> H1      H1    )
     4.83996982E-04    2          25        35   # BR(H3 -> H1      H2    )
     4.05475613E-05    2          35        35   # BR(H3 -> H2      H2    )
#
#               PDG    Width QCD and EW
DECAY QCD&EW    38     1.42137962E+01   # H3 decays with QCD and EW correctis
                       7                # Renormalization Scheme Number
       0.90516083E+00   # Corresponding mixing angle   alpha1
      -0.15530063E+00   # Corresponding mixing angle   alpha2
      -0.15445673E+01   # Corresponding mixing angle   alpha3
       0.13029492E+01   # Corresponding tan(beta)
       0.15169628E+06   # Corresponding m_12^2
       0.24408400E+04   # Corresponding v_S
#          BR         NDA      ID1       ID2
     9.71729023E-04    2           5        -5   # BR(H3 -> b       bb      )
     1.43688626E-04    2         -15        15   # BR(H3 -> tau+    tau -   )
     4.96243881E-07    2         -13        13   # BR(H3 -> mu+     mu-     )
     3.68118807E-07    2           3        -3   # BR(H3 -> s       sb      )
     1.82112984E-05    2           4        -4   # BR(H3 -> c       cb      )
     9.81534840E-01    2           6        -6   # BR(H3 -> t       tb      )
     2.39561238E-03    2          21        21   # BR(H3 -> g       g       )
     6.93146781E-06    2          22        22   # BR(H3 -> gam     gam     )
     1.93043925E-06    2          22        23   # BR(H3 -> Z       gam     )
     2.36691905E-03    2          24       -24   # BR(H3 -> W+      W-      )
     6.73441154E-04    2          23        23   # BR(H3 -> Z       Z       )
     6.46052130E-03    2          25        25   # BR(H3 -> H1      H1      )
     5.11304746E-03    2          25        35   # BR(H3 -> H1      H2      )
     3.12263142E-04    2          35        35   # BR(H3 -> H2      H2      )
#            PDG    Width QCD Only
DECAY QCD    36     2.06377878E+01   # A decays with QCD corrections only
                       7                # Renormalization Scheme Number
       0.90516083E+00   # Corresponding mixing angle   alpha1
      -0.15530063E+00   # Corresponding mixing angle   alpha2
      -0.15445673E+01   # Corresponding mixing angle   alpha3
       0.13029492E+01   # Corresponding tan(beta)
       0.14967045E+06   # Corresponding m_12^2
       0.24408400E+04   # Corresponding v_S
#          BR         NDA      ID1       ID2
     8.54499665E-04    2           5        -5   # BR(A -> b       bb       )
     1.26183976E-04    2         -15        15   # BR(A -> tau+    tau  -   )
     4.46200865E-07    2         -13        13   # BR(A -> mu+     mu-      )
     3.21263372E-07    2           3        -3   # BR(A -> s       sb       )
     1.46395489E-05    2           4        -4   # BR(A -> c       cb       )
     9.93049035E-01    2           6        -6   # BR(A -> t       tb       )
     2.49656795E-03    2          21        21   # BR(A -> g       g        )
     7.27483596E-06    2          22        22   # BR(A -> gam     gam      )
     2.39953028E-06    2          22        23   # BR(A -> Z       gam      )
     6.31266102E-04    2          23        25   # BR(A -> Z       H1       )
     2.64866795E-03    2          23        35   # BR(A -> Z       H2       )
     2.37168383E-04    2          23        38   # BR(A -> Z       H3       )
#
#               PDG    Width QCD and EW
DECAY QCD&EW    36     1.97238031E+01   # A decays with QCD and EW corrections
                       7                # Renormalization Scheme Number
       0.90516083E+00   # Corresponding mixing angle   alpha1
      -0.15530063E+00   # Corresponding mixing angle   alpha2
      -0.15445673E+01   # Corresponding mixing angle   alpha3
       0.13029492E+01   # Corresponding tan(beta)
       0.14967045E+06   # Corresponding m_12^2
       0.24408400E+04   # Corresponding v_S
#          BR         NDA      ID1       ID2
     7.80198466E-04    2           5        -5   # BR(A -> b       bb       )
     1.15123849E-04    2         -15        15   # BR(A -> tau+    tau  -   )
     3.97333090E-07    2         -13        13   # BR(A -> mu+     mu-      )
     3.02897142E-07    2           3        -3   # BR(A -> s       sb       )
     1.36486632E-05    2           4        -4   # BR(A -> c       cb       )
     9.92644607E-01    2           6        -6   # BR(A -> t       tb       )
     2.61225684E-03    2          21        21   # BR(A -> g       g        )
     7.61194583E-06    2          22        22   # BR(A -> gam     gam      )
     2.51072252E-06    2          22        23   # BR(A -> Z       gam      )
     9.68311638E-04    2          23        25   # BR(A -> Z       H1      )
     2.67851520E-03    2          23        35   # BR(A -> Z       H2      )
     1.76515248E-04    2          23        38   # BR(A -> Z       H3      )
#            PDG    Width QCD Only
DECAY QCD    37     2.17468065E+01   # H+ decays with QCD corrections only
                       7                # Renormalization Scheme Number
       0.90516083E+00   # Corresponding mixing angle   alpha1
      -0.15530063E+00   # Corresponding mixing angle   alpha2
      -0.15445673E+01   # Corresponding mixing angle   alpha3
       0.13029492E+01   # Corresponding tan(beta)
       0.14872861E+06   # Corresponding m_12^2
       0.24408400E+04   # Corresponding v_S
#          BR         NDA      ID1       ID2
     1.41706224E-06    2           4        -5   # BR(H+ -> c       bb     )
     1.25872256E-04    2         -15        16   # BR(H+ -> tau+    nu_tau )
     4.45098078E-07    2         -13        14   # BR(H+ -> mu+     nu_mu  )
     1.01136265E-08    2           2        -5   # BR(H+ -> u       bb     )
     1.54202891E-08    2           2        -3   # BR(H+ -> u       sb     )
     7.02521008E-07    2           4        -1   # BR(H+ -> c       db     )
     1.34140274E-05    2           4        -3   # BR(H+ -> c       sb     )
     9.51944776E-01    2           6        -5   # BR(H+ -> t       bb     )
     1.55493512E-03    2           6        -3   # BR(H+ -> t       sb     )
     7.16118317E-05    2           6        -1   # BR(H+ -> t       db     )
     7.13310769E-04    2          24        25   # BR(H+ -> W+      H1     )
     3.15960499E-03    2          24        35   # BR(H+ -> W+      H2     )
     4.24092631E-02    2          24        38   # BR(H+ -> W+      H3     )
     4.62195001E-06    2          24        36   # BR(H+ -> W+      A      )
#
#               PDG    Width QCD and EW
DECAY QCD&EW    37     1.99772998E+01   # H+ decays with QCD and EW corrections
                       7                # Renormalization Scheme Number
       0.90516083E+00   # Corresponding mixing angle   alpha1
      -0.15530063E+00   # Corresponding mixing angle   alpha2
      -0.15445673E+01   # Corresponding mixing angle   alpha3
       0.13029492E+01   # Corresponding tan(beta)
       0.14872861E+06   # Corresponding m_12^2
       0.24408400E+04   # Corresponding v_S
#          BR         NDA      ID1       ID2
     1.27362693E-06    2           4        -5   # BR(H+ -> c       bb     )
     1.14241461E-04    2         -15        16   # BR(H+ -> tau+    nu_tau )
     3.93843731E-07    2         -13        14   # BR(H+ -> mu+     nu_mu  )
     8.01271883E-09    2           2        -5   # BR(H+ -> u       bb     )
     1.45819891E-08    2           2        -3   # BR(H+ -> u       sb     )
     6.65248800E-07    2           4        -1   # BR(H+ -> c       db     )
     1.27376814E-05    2           4        -3   # BR(H+ -> c       sb     )
     9.50944592E-01    2           6        -5   # BR(H+ -> t       bb     )
     1.55565603E-03    2           6        -3   # BR(H+ -> t       sb     )
     7.13525053E-05    2           6        -1   # BR(H+ -> t       db     )
     7.79767060E-04    2          24        25   # BR(H+ -> W+      H1     )
     3.14065971E-03    2          24        35   # BR(H+ -> W+      H2     )
     4.33736066E-02    2          24        38   # BR(H+ -> W+      H3     )
     5.03134325E-06    2          24        36   # BR(H+ -> W+      A      )
\end{lstlisting}

As can be inferred from the file, the branching ratios of
the lightest CP-even Higgs boson $H_1$ are SM-like. 
When these are compared with those generated by the code {\tt
  N2HDECAY} \cite{Muhlleitner:2016mzt}, there are differences due to
the rescaling factor $G_F^\text{calc}/G_F =1.025366$, which is applied
in {\texttt{ewN2HDECAY}} to consistently  combine the EW-corrected
decay widths with the decay widths generated by
{\texttt{N2HDECAY}}. Additional differences arise  in the 
decay widths into massive vector bosons $V=W,Z$, of around
2\%. This is because {\texttt{N2HDECAY}} throughout computes these
decay widths using the double off-shell formula while
{\texttt{ewN2HDECAY}} uses the on-shell formula for Higgs boson masses
above the threshold. 

\subsection{Exemplary Output File for the Electroweak Partial Decay Widths} 
The exemplary output file {\texttt{ewn2hdecay\_EW.out}} contains the LO
and electroweak NLO partial decay widths. The content of the file is
presented in the following. 
\begin{lstlisting}
#
BLOCK DCINFO  # Decay Program information
     1   ewN2HDECAY  # decay calculator
     2   1.0.0       # version number
#
BLOCK SMINPUTS  # Standard Model inputs
         2     1.19596488E-05   # G_F [GeV^-2]
         3     1.18000000E-01   # alpha_S(M_Z)^MSbar
         4     9.11876000E+01   # M_Z on-shell mass
         5     4.18000000E+00   # mb(mb)^MSbar
         6     1.73200000E+02   # mt pole mass
         7     1.77682000E+00   # mtau pole mass
         8     4.84141297E+00   # mb pole mass
         9     1.43141297E+00   # mc pole mass
        10     1.05658372E-01   # muon mass
        11     8.03850000E+01   # M_W on-shell mass
        12     2.08500000E+00   # W boson total width
        13     2.49520000E+00   # Z boson total width
#
BLOCK N2HDMINPUTS  # N2HDM inputs of reference scheme
         1                  2   # N2HDM type
         2     1.32331000E+00   # tan(beta)
         3     1.82778000E+05   # M_12^2
         4     1.25090000E+02   # M_H1
         5     2.86094000E+02   # M_H2
         6     6.48564000E+02   # M_H3
         7     7.22000000E+02   # M_A
         8     7.58918000E+02   # M_CH
         9     9.09079000E-01   # alpha1
        10    -1.55397000E-01   # alpha2
        11    -1.54459000E+00   # alpha3
        12     2.44084000E+03   # v_S
        13                  7   # renormalization scheme EW corrs
        14                  5   # reference ren. scheme EW corrs
        15     125.09D0         # ren. scale of MSbar parameters
        16     MIN              # ren. scale at which decay is evaluated
#
BLOCK VCKMIN  # CKM mixing
         1     9.99100000E-01   # V_tb
         2     4.04000000E-02   # V_ts
         3     8.67000000E-03   # V_td
         4     4.12000000E-02   # V_cb
         5     9.73440000E-01   # V_cs
         6     2.25200000E-01   # V_cd
         7     3.51000000E-03   # V_ub
         8     2.25340000E-01   # V_us
         9     9.74270000E-01   # V_ud
#
#                 PDG
LO DECAY WIDTH    25   # H1 non-zero LO EW decay widths of on-shell and non-loop induced decays
                    7   # Renormalization Scheme Number
       0.90516083E+00   # Corresponding mixing angle   alpha1
      -0.15530063E+00   # Corresponding mixing angle   alpha2
      -0.15445673E+01   # Corresponding mixing angle   alpha3
       0.13029492E+01   # Corresponding tan(beta)
       0.18277800E+06   # Corresponding m_12^2
       0.24408400E+04   # Corresponding v_S
#         WIDTH       NDA         ID1       ID2
     5.89110455E-03    2           5        -5   # GAM(H1 -> b       bb )
     2.66567492E-04    2         -15        15   # GAM(H1 -> tau+    tau)
     9.43739741E-07    2         -13        13   # GAM(H1 -> mu+     mu-)
     2.28882843E-06    2           3        -3   # GAM(H1 -> s       sb )
     4.96099310E-04    2           4        -4   # GAM(H1 -> c       cb )
#
#                 PDG
NLO DECAY WIDTH    25   # H1 non-zero NLO EW decay widths of on-shell and non-loop induced decays
                    7   # Renormalization Scheme Number
       0.90516083E+00   # Corresponding mixing angle   alpha1
      -0.15530063E+00   # Corresponding mixing angle   alpha2
      -0.15445673E+01   # Corresponding mixing angle   alpha3
       0.13029492E+01   # Corresponding tan(beta)
       0.18277800E+06   # Corresponding m_12^2
       0.24408400E+04   # Corresponding v_S
#         WIDTH       NDA         ID1       ID2
     5.69902477E-03    2           5        -5   # GAM(H1 -> b       bb    )
     2.54863852E-04    2         -15        15   # GAM(H1 -> tau+    tau-  )
     8.82552598E-07    2         -13        13   # GAM(H1 -> mu+     mu-   )
     2.23273559E-06    2           3        -3   # GAM(H1 -> s       sb    )
     4.76534904E-04    2           4        -4   # GAM(H1 -> c       cb    )
#
#                 PDG
LO DECAY WIDTH    35   # H2 non-zero LO EW decay widths of on-shell and non-loop induced decays
                    7   # Renormalization Scheme Number
       0.90516083E+00   # Corresponding mixing angle   alpha1
      -0.15530063E+00   # Corresponding mixing angle   alpha2
      -0.15445673E+01   # Corresponding mixing angle   alpha3
       0.13029492E+01   # Corresponding tan(beta)
       0.16715360E+06   # Corresponding m_12^2
       0.24408400E+04   # Corresponding v_S
#         WIDTH       NDA         ID1       ID2
     4.91587719E-04    2           5        -5   # GAM(H2 -> b       bb )
     2.21038791E-05    2         -15        15   # GAM(H2 -> tau+    tau)
     7.81789109E-08    2         -13        13   # GAM(H2 -> mu+     mu-)
     1.89605235E-07    2           3        -3   # GAM(H2 -> s       sb )
     2.08960570E-05    2           4        -4   # GAM(H2 -> c       cb )
     1.17815813E-01    2          24       -24   # GAM(H2 -> W+      W- )
     5.18688369E-02    2          23        23   # GAM(H2 -> Z       Z  )
     8.19267433E-02    2          25        25   # GAM(H2 -> H1      H1 )
#
#                 PDG
NLO DECAY WIDTH    35   # H2 non-zero NLO EW decay widths of on-shell and non-loop induced decays
                    7   # Renormalization Scheme Number
       0.90516083E+00   # Corresponding mixing angle   alpha1
      -0.15530063E+00   # Corresponding mixing angle   alpha2
      -0.15445673E+01   # Corresponding mixing angle   alpha3
       0.13029492E+01   # Corresponding tan(beta)
       0.16715360E+06   # Corresponding m_12^2
       0.24408400E+04   # Corresponding v_S
#         WIDTH       NDA         ID1       ID2
     4.61989652E-04    2           5        -5   # GAM(H2 -> b       bb    )
     2.06423964E-05    2         -15        15   # GAM(H2 -> tau+    tau-  )
     7.13757478E-08    2         -13        13   # GAM(H2 -> mu+     mu-   )
     1.81751713E-07    2           3        -3   # GAM(H2 -> s       sb    )
     2.00090392E-05    2           4        -4   # GAM(H2 -> c       cb    )
     1.18674571E-01    2          24       -24   # GAM(H2 -> W+      W-    )
     5.20176206E-02    2          23        23   # GAM(H2 -> Z       Z     )
     1.16353468E-01    2          25        25   # GAM(H2 -> H1      H1    )
#
#                 PDG
LO DECAY WIDTH    38   # H3 non-zero LO EW decay widths of on-shell and non-loop induced decays
                    7   # Renormalization Scheme Number
       0.90516083E+00   # Corresponding mixing angle   alpha1
      -0.15530063E+00   # Corresponding mixing angle   alpha2
      -0.15445673E+01   # Corresponding mixing angle   alpha3
       0.13029492E+01   # Corresponding tan(beta)
       0.15169628E+06   # Corresponding m_12^2
       0.24408400E+04   # Corresponding v_S
#         WIDTH       NDA         ID1       ID2
     5.08405647E-02    2           5        -5   # GAM(H3 -> b       bb )
     2.28327056E-03    2         -15        15   # GAM(H3 -> tau+    tau)
     8.07416876E-06    2         -13        13   # GAM(H3 -> mu+     mu-)
     1.95820644E-05    2           3        -3   # GAM(H3 -> s       sb )
     1.64155530E-03    2           4        -4   # GAM(H3 -> c       cb )
     1.45228111E+01    2           6        -6   # GAM(H3 -> t       tb )
     1.90555811E-02    2          24       -24   # GAM(H3 -> W+      W- )
     9.27982692E-03    2          23        23   # GAM(H3 -> Z       Z  )
     2.64224042E-02    2          25        25   # GAM(H3 -> H1      H1 )
     7.10491271E-03    2          25        35   # GAM(H3 -> H1      H2 )
     5.95224546E-04    2          35        35   # GAM(H3 -> H2      H2 )
#
#                 PDG
NLO DECAY WIDTH    35   # H3 non-zero NLO EW decay widths of on-shell and non-loop induced decays
                    7   # Renormalization Scheme Number
       0.90516083E+00   # Corresponding mixing angle   alpha1
      -0.15530063E+00   # Corresponding mixing angle   alpha2
      -0.15445673E+01   # Corresponding mixing angle   alpha3
       0.13029492E+01   # Corresponding tan(beta)
       0.15169628E+06   # Corresponding m_12^2
       0.24408400E+04   # Corresponding v_S
#         WIDTH       NDA         ID1       ID2
     4.53131283E-02    2           5        -5   # GAM(H3 -> b       bb    )
     2.04236084E-03    2         -15        15   # GAM(H3 -> tau+    tau-  )
     7.05350937E-06    2         -13        13   # GAM(H3 -> mu+     mu-   )
     1.80796084E-05    2           3        -3   # GAM(H3 -> s       sb    )
     1.49442795E-03    2           4        -4   # GAM(H3 -> c       cb    )
     1.39109721E+01    2           6        -6   # GAM(H3 -> t       tb    )
     3.36429049E-02    2          24       -24   # GAM(H3 -> W+      W-    )
     9.57215530E-03    2          23        23   # GAM(H3 -> Z       Z     )
     9.18285329E-02    2          25        25   # GAM(H3 -> H1      H1    )
     7.26758144E-02    2          25        35   # GAM(H3 -> H1      H2    )
     4.43844464E-03    2          35        35   # GAM(H3 -> H2      H2    )
#
#                 PDG
LO DECAY WIDTH    36   # A non-zero LO EW decay widths of on-shell and non-loop induced decays
                    7   # Renormalization Scheme Number
       0.90516083E+00   # Corresponding mixing angle   alpha1
      -0.15530063E+00   # Corresponding mixing angle   alpha2
      -0.15445673E+01   # Corresponding mixing angle   alpha3
       0.13029492E+01   # Corresponding tan(beta)
       0.14967045E+06   # Corresponding m_12^2
       0.24408400E+04   # Corresponding v_S
#         WIDTH       NDA         ID1       ID2
     5.79978912E-02    2           5        -5   # GAM(A -> b       bb    )
     2.60415812E-03    2         -15        15   # GAM(A -> tau+    tau-  )
     9.20859876E-06    2         -13        13   # GAM(A -> mu+     mu-   )
     2.23333665E-05    2           3        -3   # GAM(A -> s       sb    )
     1.75923370E-03    2           4        -4   # GAM(A -> c       cb    )
     2.25987982E+01    2           6        -6   # GAM(A -> t       tb    )
     1.30279359E-02    2          23        25   # GAM(A -> Z       H1    )
     5.46626471E-02    2          23        35   # GAM(A -> Z       H2    )
#
#                 PDG
NLO DECAY WIDTH    36   # A non-zero NLO EW decay widths of on-shell and non-loop induced decays
                    7   # Renormalization Scheme Number
       0.90516083E+00   # Corresponding mixing angle   alpha1
      -0.15530063E+00   # Corresponding mixing angle   alpha2
      -0.15445673E+01   # Corresponding mixing angle   alpha3
       0.13029492E+01   # Corresponding tan(beta)
       0.14967045E+06   # Corresponding m_12^2
       0.24408400E+04   # Corresponding v_S
#         WIDTH       NDA         ID1       ID2
     5.06096011E-02    2           5        -5   # GAM(A -> b       bb    )
     2.27068012E-03    2         -15        15   # GAM(A -> tau+    tau-  )
     7.83691964E-06    2         -13        13   # GAM(A -> mu+     mu-   )
     2.01240631E-05    2           3        -3   # GAM(A -> s       sb    )
     1.56752139E-03    2           4        -4   # GAM(A -> c       cb    )
     2.15891704E+01    2           6        -6   # GAM(A -> t       tb    )
     1.90987881E-02    2          23        25   # GAM(A -> Z       H1    )
     5.28305065E-02    2          23        35   # GAM(A -> Z       H2    )
#
#                 PDG
LO DECAY WIDTH    37   # H+ non-zero LO EW decay widths of on-shell and non-loop induced decays
                    7   # Renormalization Scheme Number
       0.90516083E+00   # Corresponding mixing angle   alpha1
      -0.15530063E+00   # Corresponding mixing angle   alpha2
      -0.15445673E+01   # Corresponding mixing angle   alpha3
       0.13029492E+01   # Corresponding tan(beta)
       0.14872861E+06   # Corresponding m_12^2
       0.24408400E+04   # Corresponding v_S
#         WIDTH       NDA         ID1       ID2
     1.06619789E-04    2           4        -5   # GAM(H+ -> c       bb   )
     2.73731960E-03    2         -15        16   # GAM(H+ -> tau+    nu_tau)
     9.67946175E-06    2         -13        14   # GAM(H+ -> mu+     nu_mu)
     7.51082775E-07    2           2        -5   # GAM(H+ -> u       bb   )
     1.19203342E-06    2           2        -3   # GAM(H+ -> u       sb   )
     9.37817403E-05    2           4        -1   # GAM(H+ -> c       db   )
     1.77450991E-03    2           4        -3   # GAM(H+ -> c       sb   )
     2.43287119E+01    2           6        -5   # GAM(H+ -> t       bb   )
     3.97057154E-02    2           6        -3   # GAM(H+ -> t       sb   )
     1.82863984E-03    2           6        -1   # GAM(H+ -> t       db   )
     1.55122312E-02    2          24        25   # GAM(H+ -> W+      H1   )
     6.87113183E-02    2          24        35   # GAM(H+ -> W+      H2   )
     9.22266037E-01    2          24        38   # GAM(H+ -> W+      H3   )
#
#                 PDG
NLO DECAY WIDTH    37   # H+ non-zero NLO EW decay widths of on-shell and non-loop induced decays
                    7   # Renormalization Scheme Number
       0.90516083E+00   # Corresponding mixing angle   alpha1
      -0.15530063E+00   # Corresponding mixing angle   alpha2
      -0.15445673E+01   # Corresponding mixing angle   alpha3
       0.13029492E+01   # Corresponding tan(beta)
       0.14872861E+06   # Corresponding m_12^2
       0.24408400E+04   # Corresponding v_S
#         WIDTH       NDA         ID1       ID2
     8.80303489E-05    2           4        -5   # GAM(H+ -> c       bb   )
     2.28223592E-03    2         -15        16   # GAM(H+ -> tau+    nu_tau)
     7.86793427E-06    2         -13        14   # GAM(H+ -> mu+     nu_mu)
     5.46640859E-07    2           2        -5   # GAM(H+ -> u       bb   )
     1.03550925E-06    2           2        -3   # GAM(H+ -> u       sb   )
     8.15801254E-05    2           4        -1   # GAM(H+ -> c       db   )
     1.54792859E-03    2           4        -3   # GAM(H+ -> c       sb   )
     2.23256376E+01    2           6        -5   # GAM(H+ -> t       bb   )
     3.64918286E-02    2           6        -3   # GAM(H+ -> t       sb   )
     1.67376283E-03    2           6        -1   # GAM(H+ -> t       db   )
     1.55776403E-02    2          24        25   # GAM(H+ -> W+      H1   )
     6.27419005E-02    2          24        35   # GAM(H+ -> W+      H2   )
     8.66487541E-01    2          24        38   # GAM(H+ -> W+      H3   )
\end{lstlisting} 

As can be read off from the output file, the EW corrections reduce the
decay widths of the SM-like Higgs boson $H_1$. The relative NLO EW
corrections, $\Delta^{\text{EW}} = (\Gamma^{\text{EW}} -
  \Gamma^{\text{LO}})/\Gamma^{\text{LO}}$, range between -6.5 and
  -2.5\% for the decays $\Gamma(H_1 \to \mu^+\mu^-)$ and $\Gamma(H_1 \to
  s\bar{s})$, respectively. For the heavier Higgs bosons the NLO EW corrections can
  be considerably larger. We remind the reader also
  that only LO and NLO EW-corrected decay widths are given out for
  on-shell and non-loop induced decays.

\end{appendix}


\end{document}